\documentstyle[aps,pre,12pt]{revtex}

\begin{document}
\title{Exact solutions to chaotic and stochastic systems}
\author{J. A. Gonz\'{a}lez$^{1}$, L. I. Reyes$^{1,2}$, and L. E. Guerrero$^{2}$}
\address{$^{1}$Centro de F\'{\i }sica, Instituto Venezolano de Investigaciones\\
Cient\'{\i }ficas, Apartado Postal 21827, Caracas 1020-A, Venezuela}
\address{$^{2}$Departamento de F\'{\i }sica, Universidad Sim\'{o}n Bol\'{\i }var,\\
Apartado 89000, Caracas 1080-A, Venezuela}
\date{To appear in \textit{Chaos}, March 2001}
\maketitle

\begin{abstract}
\noindent

We investigate functions that are exact solutions to chaotic dynamical
systems. A generalization of these functions can produce truly random
numbers. For the first time, we present solutions to random maps. This
allows us to check, analytically, some recent results about the complexity
of random dynamical systems. We confirm the result that a negative Lyapunov
exponent does not imply predictability in random systems. We test the
effectiveness of forecasting methods in distinguishing between chaotic and
random time series. Using the explicit random functions, we can give
explicit analytical formulas for the output signal in some systems with
stochastic resonance. We study the influence of chaos on the stochastic
resonance. We show, theoretically, the existence of a new type of solitonic
stochastic resonance, where the shape of the kink is crucial. Using our
models we can predict specific patterns in the output signal of stochastic
resonance systems.
\end{abstract}


%\twocolumn[\hsize\textwidth\columnwidth\hsize\csname @twocolumnfalse\endcsname

%\widetext

\preprint{HEP/123-qed} %]

%\narrowtext

\newpage

{\bf Recently, many outstanding papers \cite{Ferrenberg,D'Souza,Nogues,James}
have stated the importance of having true random models. The best existing
pseudorandom number generators can yield incorrect results due to the
``unavoidable'' correlations that appear between the generated values \cite
{Ferrenberg,D'Souza,Nogues}. On the other hand, there is a great interest in
random dynamical systems \cite{Yu,Paladin,Loreto}. In recent years, there
has been much discussion about the transition to chaos and the way to
characterize predictability and complexity in these systems \cite
{Paladin,Loreto}. There is also strong controversy about the existing
methods to distinguish chaotic and completely random systems \cite
{Sugihara,Wales,Tsonis}.}
{\bf In the present paper we investigate explicit functions that are exact
solutions to nonlinear chaotic maps. A generalization of these functions can
produce truly random sequences. Even if the initial conditions are known
exactly, the next values are in principle unpredictable from the previous
values. These functions cannot be expressed as a map of type }$%
X_{n+1}=g\left( X_{n},X_{n-1},...,X_{n-r+1}\right) ${\bf . Using some of
these functions we can exactly solve random maps as the following:}

\begin{equation}
X_{n+1}=f\left( X_{n},I_{n}\right),  \label{Eq1}
\end{equation}
{\bf where }$I_{n}${\bf \ is a random variable.}
{\bf We can confirm the result \cite{Paladin,Loreto} that a negative
Lyapunov exponent does not imply predictability in random systems. We show
that the forecasting methods \cite{Sugihara,Wales,Tsonis} are very effective
in distinguishing chaos from random time series. We investigate the influence
of the level of chaos on the stochastic resonance (SR). We can give explicit
analytical formulas for the output signal of some systems with stochastic
resonance. We show the existence of a new type of solitonic stochastic
resonance (SSR), where the shape of the kink is crucial.}

\section{Introduction}

There is the common belief that, as truly random numbers should be
unpredictable in advance, they must be produced by random physical processes
such as radioactive decay, thermal noise in electronic devices, cosmic ray
arrival time, etc. \cite{James}.

Knowing the past and present values should give no information as to future
outcomes of a truly random variable \cite{D'Souza}. Thus, a recursive
mathematical algorithm should not be able to describe a truly random
process. From this, it seems, that deterministic randomness is inherently
unattainable \cite{D'Souza}.

Here we have two problems as a motivation for our work:

\begin{enumerate}
\item  How to describe theoretically these physical phenomena that are truly
random.

\item  How to produce truly random numbers, which are necessary in different
physical calculations such as Monte Carlo method.
\end{enumerate}

The purpose of our study is to find explicit functions that produce truly
random dynamics. These functions can be used as random number generators and
as analytical solutions to nonlinear random systems.

It is well known \cite{Ulam,Stein} that the function $X_{n}=\sin ^{2}\left(
\theta \pi 2^{n}\right) $ is the general solution to the logistic map $%
X_{n+1}=4X_{n}\left( 1-X_{n}\right) $. Recently, other chaotic maps have
been reported to have exact solutions \cite
{Katsura,Kawamoto,Brown,Umeno,González,Nazareno,González2}. In the present
paper we will investigate in detail a generalization of the solution to the
logistic map:

\begin{equation}
X_{n}=\sin ^{2}\left( \theta \pi z^{n}\right),  \label{Eq2}
\end{equation}
where $z$ is a real number.

For $z$ integer, function (\ref{Eq2}) is the general solution to the
family of maps:

\begin{equation}
X_{n+1}=\sin ^{2}\left( z\arcsin \sqrt{X_{n}}\right).  \label{Eq3}
\end{equation}

Even for a real $z$ we can calculate the Lyapunov exponent of map (\ref
{Eq3}) exactly: $\lambda =\ln z$.

For $z>1$, map (\ref{Eq3}) is chaotic. Nevertheless, for fractionary $z$
the dynamics contained in function (\ref{Eq2}) is quite different from that
of map (\ref{Eq3}). In fact, for a fractionary $z$, the first-return map
generated by Eq. (\ref{Eq2}) is multivalued (see Figs. 1 and 2). Let $z$ be
a rational number expressed as $z=p/q$, where $p$ and $q$ are relative prime
numbers. Then the first-return map produced by function (\ref{Eq2}) is a
curve such that, in general, for a value of $X_{n}$ we will have $q$ values
of $X_{n+1}$. On the other hand, for a value of $X_{n+1}$ we will have $p$
values of $X_{n}$. Geometrically, these curves are Lissajous figures \cite
{González}. But we should note that their meaning here is very different
from that in their original definition. In this context, they represent
chaotic first-return maps. For $z$ irrational, the first-return map is a
random set of points as shown in Fig. 3.

The paper is organized as follows. In Sec. II we study the properties of
the functions $X_{n}=\sin ^{2}\left( \theta \pi z^{n}\right)$. We present a
rigorous proof that, for $z$ fractionary, the produced sequences are
absolutely unpredictable in advance. Moreover, the outcomes are completely
independent. In Sec. III we discuss the use of these functions in actual
numerical calculations. Section IV is dedicated to random maps of type $%
X_{n+1}=f(X_n,I_n)$, where $I_n$ is a random variable. Function (\ref
{Eq2}) can help one to find analytical solutions to these maps. We calculate
exactly the complexity of a random map. This allows us to check some recent
results about the complexity and predictability of random maps. In Sec. V
we address the problem of distinguishing chaos from random time series. For
this, we check the effectivity of the so-called ``nonlinear forecasting
methods.'' Section VI is devoted to stochastic resonance (SR). First we give
some introductory remarks about the historical developments in SR.
Considering the fact that we can calculate exactly the Lyapunov exponent of
a class of chaotic maps, we are able to investigate the influence of the
level of chaos on SR. This is done first in the most common setup for SR: a
bistable system. Then, we investigate the so-called nonlinear static systems
with SR. For these systems, we can present explicit analytical functions
that describe the output of the system. Using the functions we can
investigate the actual dynamics of the system. Finally, based on theoretical
investigations, we show the existence of a new type of solitonic stochastic
resonance, where the shape of the kink is crucial.

\section{Explicit stochastic functions}

After a rigorous analysis of function (\ref{Eq2}) we arrive at interesting
conclusions. For most fractionary $z>1$ function (\ref{Eq2}) is not only
chaotic, but its next value is impossible to predict (from the previous
values) unless $\theta$ is exactly known. When $z$ is an integer, the initial
condition $X_{0}$ defines univocally the value of $\theta $ (any value of $%
\theta $ out of the interval $0<\theta <1$ defining $X_{0}$ is equivalent to
one in that interval). If $z$ is fractionary, this is not so. There exists
an infinite number of values of $\theta $ that satisfy the initial
conditions. The time series produced for different values of $\theta $
satisfying the initial conditions is different in most cases. The fact that
we know the initial conditions does not imply that we can determine $\theta $%
. So the next value is unpredictable.

Let us consider the case $z=3/2$ (see Fig. 2). If we wish to calculate $%
X_{n+1}$ from the value $X_{n}$ we will have two choices:

\begin{equation}
X_{n+1}=\frac{1}{2}\left[ 1\pm \left( 1-4X_{n}\right) \left( 1-X_{n}\right)
^{1/2}\right].  \label{Eq4}
\end{equation}

The value $X_{n+1}$ could be expressed as a well-defined function of the
previous values if $\left( 1-X_{n}\right) ^{1/2}$ could be a rational
function of the previous values. However, each time we try to do this we
meet the same difficulty because the previous values are also irrational
functions of the past values. This process can continue up to infinity.

A different way to see this phenomenon is the following. Consider the family
of functions

\begin{equation}
X_{n}^{k}=\sin ^{2}\left[ \left( \theta _{0}+k\right) \pi z^{n}\right],
\label{Eq5}
\end{equation}
where $\theta =\theta _{0}+k$, $k$ is integer.

For all $k$, the time series $X_{n}^{k}$ ($k$ fixed, $n$ as time) have the
same initial conditions. If $z$ is an integer, the initial condition defines
the complete sequence (see Table I). However, for $z$ fractionary all the
time series are different. This is because the period of function $X_{n}^{k}$
(now $n$ is fixed and $k$ is variable) is different for different $n$ (for
instance, when $z=3/2$, the period of $X_{n}^{k}$ is $2^{n}$). In general,
for $z=p/q$, the period is $q^{n}$. That is, $X_{n+1}$ cannot be determined
by $X_{n}$. Moreover, $X_{n+1}$ cannot be determined by any number of
previous values. Let us see the following example with $z=3/2$. Suppose $%
X_{n}=0$. Now we have two possibilities $X_{n+1}=0$ or $X_{n+1}=1$ (see
Table II). Assume $\theta _{0}=0$ and $n=0$. For any $\theta =k$ ($k$
integer), $X_{n}=0$. Now, $X_{n+1}=\sin ^{2}\left[ \left( 3/2\right) k\pi
\right] $. So, $X_{n+1}=0$ for $k$ even, and $X_{n+1}=1$ for $k$ odd. But
there is no way we can know $k$ from the statement $X_{n}=0$ (for all $k$
integers this statement is true). This uncertainty about the next value is
present for all points $X_{n}$ except $X_{n}=1/4$ and $X_{n}=1$. But these
two points are a set of zero measure. That is, for almost all the points in
the interval $0<X_{n}<1$, the next value is unpredictable.

For $z$ irrational there are infinite possibilities for $X_{n+1}$. All
values are unpredictable. But let us continue with the simple case $z=3/2$.
Suppose now that $\theta =2^{m}$, where $m$ is an integer. Note that in this
case $X_{0}=0$. But, unless we know $\theta $, we never will know when the
value $X_{m+1}$ will be equal to $1$ (see Table II). We can have a string of 
$m+1$ zeros ($m$ can be as large as we wish) and only in the point $X_{m+1}$
does the sequence change from a string of zeros to the value $1$. So, for any
finite number $m+1$ of previous values $X_{0}$, $X_{1}$, $X_{2}$,..., $X_{m}$%
; the next value is not defined by the previous values. Note that in this
example we can have a string of zeros, but this is because the value $%
X_{n}=0 $ is a pseudofixed point of the map $\left( X_{n},X_{n+1}\right) $
due to the intersection of the graph in Fig. 2(a) with the line $X_{n+1}=X_{n}$%
. However, in general, the sequence is very stochastic. On the other hand,
the uncertainty about which is the next value remains for all the points in
the interval $0\leq X_{n}\leq 1$ except for $X_{n}=1/4$ and $X_{n}=1$. The
general uncertainty increases for $p>q>2$ (see Table III). In this case, the
unpredictability is true for all values of $X_{n}$.

On the other hand, if $z$ is irrational, then the points on the first-return
map $\left( X_{n},X_{n+1}\right) $ will fill the square $0\leq X_{n}\leq 1$; 
$0\leq X_{n+1}\leq 1$ (see Fig. 3 and Table IV). For a large but finite
number $n$, the map is an erratic set of points (we should exclude the
numbers of type $z=m^{1/k}$, where $m$ and $k$ are integers, because in this
case the sequence is predictable given $k$ previous values).

Note that we can consider $X_{n}^{k}$ defined by Eq. (\ref{Eq5}) as an infinite
matrix, where the ``columns'' are the stochastic sequences (dependence on $n$%
) and the horizontal ``rows'' are periodic (or quasiperiodic for irrational $%
z$) sequences that represent the dependence on $k$. For $z=p/q$, the
``rows'' are periodic sequences with period $q^{n}$ (see Tables II and III).
We see that all the row sequences have different periods. So, all the
column sequences are generally different. However, for each integer $m$,
there is an infinite set of columns having a string of values of length $m$
that is identical in each number of this set. That is, in the matrix $%
X_{n}^{k}$, given an initial string of length $m=2$, we will find a string
identical to it with a period $q^{2}$. Note (in Table II) that the string ($%
0,1,1/2$) can be found in infinite columns. However, the next value is
always uncertain. It can be $X_{3}=0.1464...$ or $X_{3}=0.8536...$ . Just to
know that the previous values are ($0,1,1/2$) does not give us the knowledge
to determine the next value. The string ($%
0,1,0.5,0.8535...,0.0380...,0.9157...,0.8865...,0.0711...,0.8447...,0.9686...
$) can be found with period $2^{9}$. That is, the column number $2^{9}+1$
possesses this same string. However, the value $X_{10}$ is not always $%
X_{10}=0.7544...$ . It can be $X_{10}=0.2455...$ with the same probability.

In general, given an initial string of length $m$, we will find a string
identical to it with a period $q^{m}$. At the same time, most of these
strings possess different next values (we have seen a striking example in
the above-given text). Suppose there is a univalent function $X_{n+1}=g\left(
X_{n},X_{n-1},...,X_{n-r+1}\right) $ that is equivalent to the sequence (\ref
{Eq2}) for $z$ fractionary. If we have more than one sequence $X_{0}$, $%
X_{1} $, $X_{2}$,..., $X_{m-1}$ with different next values, then we should
decide that the map we are looking for cannot be of order $m$. If for any $m$%
, $m=1,2,3,...,\infty $; we have more than one sequence $X_{0}$, $X_{1}$, $%
X_{2} $,..., $X_{m-1}$, such that the next values are different, then such
a map does not exist.

In the above-given text we have shown that for each string of values $X_{0}$, $%
X_{1}$, $X_{2}$,..., $X_{m-1}$, there is another sequence with these same
values but with different proceeding values.

For $z$ irrational, all the row sequences are quasiperiodic and different.
The column sequences correspond to completely random sequences. These
functions can produce a set of completely independent values.

\section{Random number generators}

Now we should say some words about the use of these functions in actual
numerical calculations. The argument of function (\ref{Eq2}) increases
exponentially. So, there can be some problems in generating very large
sequences. A practical solution is to change parameters $\theta $ after a
fixed number $n=N$ of sequence values $X_{n}$. Suppose $N$ is a number for
which there are not calculation problems. For producing the new set of
values of $X_{n}$ (with a new $\theta $) we start again with $n=0$. This
procedure can be repeated the desired number of times (remember that even if
the sequence is finite, it will be unpredictable; and a sequence formed as a
set of unpredictable sequences will be also unpredictable). It can be shown
that there exists always a $\theta $ such that, with it, the original
function will produce the same sequence as that generated with the procedure
of changing $\theta $.

For the calculation of truly random numbers with function (\ref{Eq2}) the
best way is to use an irrational $z$. This irrational $z$ does not have to be a
large number. For instance, we can use $z=\pi $. The geometrical place of
the return map for $z$ irrational is the whole square $0\leq X_{n}\leq 1$; $%
0\leq X_{n+1}\leq 1$. So, we do not have to worry about the method for
determining the next value of $\theta $. For example, we can use the
following method in order to change parameter $\theta $ after each set of $N$
sequence values.

Let us define $\theta _{s}=AW_{s}$, where $s$ is the order number of $\theta 
$ in such a way that $s=1$ corresponds to the $\theta $ used for the first
set of $N$ values $X_{n}$; $s=2$ for the second set, etc.; $W_{s}$ is a
``stochastic'' sequence. For instance, the values $W_{s}$ can be obtained
from the same sequence $X_{n}$. The inequality $A>1$ should hold in order to
keep the absolute unpredictability.

Another important question about good random numbers is to have a generator
able to produce uniformly distributed points. By means of the transformation 
$Y_{n}=\left( 2/\pi \right) \arcsin \left( X_{n}^{1/2}\right) $; we can
obtain random numbers uniformly distributed on the interval $\left(
0,1\right) $ \cite{González2}. Once we have uniformly distributed random
numbers, we can use well-known transformations to generate random numbers
with any given distribution \cite{González2}.

We have performed several standard statistical tests with the functions $%
X_n=\sin^2 \left( \theta \pi z^n \right)$ (after the transformation $%
Y_{n}=\left( 2/\pi \right) \arcsin \left( X_{n}^{1/2}\right)$. Among them
are the following: the central limit theorem test, the moments calculations,
the variance calculation, and the $\chi^2$ test. The sequence $Y_n$ has
passed all these tests satisfactorily. For instance, the theoretical values
for the moments and variances are the following: $<X^n>=\frac{1}{n+1}$, $%
\sigma_n=\frac{n^2}{(2n+1)(n+1)^2}$, and these values are obtained when we
use the sequence $Y_n$.

The autocorrelation function $C_m=<Y_iY_{i+m}>-<Y_i>^2$ (where $< >$ is
the average overall $i$ with $i=1,2,3, ...$) can be shown to be zero even
for $m=1$. For the known chaotic maps (which sometimes are used as
pseudorandom number generators) $|C_m|$ decays with $m$, but there is a range
of this dependence that is related to the correlation or memory time.

Recently \cite{Pincus} a new method has been developed, which allows us to
compare the randomness of different sequences. In these works, a measure of
randomness (we will call it $R$) is introduced.

Suppose we have a sequence of values $U_1, U_2, U_3, ... , U_n$. Form a
sequence of vectors

\begin{equation}
X_{(i)}=\left[ U_i, U_{i+1}, ... , U_{i+m-1} \right].
\end{equation}

Now, we will define some variables:

\begin{equation}
C_i^m(r)=\frac{number\hspace{1mm} of\hspace{1mm} j \hspace{1mm}such \hspace{%
1mm}that\hspace{1mm} d\left[ X_{(i)},X_{(j)}\right] \le r}{N-m+1},
\end{equation}
where $d\left[ X_{(i)},X_{(j)}\right]$ is the distance between two vectors,
which is defined as follows:

\begin{equation}
d\left[ X_{(i)},X_{(j)}\right]=max \left( |U_{i+k-1} - U_{j+k-1}| \right)
\end{equation}
with $k=1,2, ... , m$.

Now we can define the measure of randomness:

\begin{equation}
R(m,r,N)=\phi_{(r)}^m-\phi_{(r)}^{m+1},
\end{equation}
where 
\begin{equation}
\phi_{(r)}^m=\frac{1}{N-m+1}\sum_{i=1}^{N-m+1}\ln C_i^m(r).
\end{equation}

This measure depends on the resolution parameter $r$ and an ``embedding''
parameter $m$. This technique has been proved to be very effective in
determining system randomness \cite{Pincus}.

For given $r$ and $m$ we have a maximum possible randomness. A sequence with
maximum randomness is uncorrelated. The randomness of our sequences $Y_n$
with $z$ irrational is the maximum possible for the given $r$ and $m$. For
instance, if $r=0.025$, the maximum possible randomness is $R=\ln 40$. The
randomness of function (\ref{Eq2}) with $z=\pi$ approaches the value $%
R=3.688 $ for increasing $N$. For comparison, the randomness of the logistic
map at the point of full chaos is $R=0.693$. Even if we further decrease $r$
and increase $m$ and $N$, for the logistic map and other usually chaotic
maps, $R$ saturates and remains constant.

On the other hand, for $r \to 0$, the randomness of function (\ref{Eq2})
with $z=\pi$ tends to the maximum possible value, i.e., $R \to \ln (1/r)$.
For $r \to 0$, it never saturates.

We should say that the pseudorandom number generators described in Ref. 
1 can pass some of the statistical tests devised to check
pseudorandomness \cite{James}. However, hidden errors in these generators
have been found \cite{Ferrenberg}. Several researchers have traced the
errors to the dependence in the pseudorandom numbers. Indeed, they are all
based on recursive algorithms.

Recently simulations of different physical systems have become the ``new
tests'' for pseudorandom number generators. Among these systems are the
following: the two-dimensional Ising model, ballistic deposition, and random
walks. Nogu\'es {\it et al} \cite{Nogues} have found that using common
pseudorandom number generators, the produced random walks present
symmetries, meaning that the generated numbers are not independent. On the
other hand, the logarithmic plot of the mean distance versus the number of
steps $N$ is not a straight line after $N > 10^5$ (in fact, it is a rapidly
decaying function).

D'Souza \textit{et al.} \cite{D'Souza} use ballistic deposition to test
the randomness of pseudorandom number generators. They found correlations in
the pseudorandom numbers and strong coupling between the model and the
generator (even generators that pass extensive statistical tests).

In a ballistic deposition model of growth, free particles initiated at
random positions above a one-dimensional substrate, descend ballistically
and stick upon first touching the surface of the growing cluster. The
substrate of length $L$ consists of discrete columns indexed by integer
values of $x$ with $1\le x \le L$. The growth interface is defined by the
maximum occupied site along each column $h(x,L)$, where $h(x,L)$ also takes
on discrete values.

The width of the growth interface $\xi_L(t)$ on average increases following
a power law behavior until reaching a steady asymptotic value, the magnitude
of which depends on the underlying substrate size $L$. For $\xi_L(t)$ we
have 
\begin{equation}
\xi_L^2(t)=\frac{1}{L}\sum_{x=1}^L\left[ h(x,t)- <h(t)>\right]^2,
\end{equation}
where $<h(t)>$ is the mean height of the surface at time $t$.

One consequence of the Kardar-Parisi-Zhang theory is that the steady state
behavior for the interface fluctuations in one dimension should resemble a
random walk, i.e., $\xi_L(t \to \infty) \sim L^{1/2}$. Thus, a random walk
again serves as a good test for random numbers. These two papers \cite
{D'Souza,Nogues} have been cited by Fisher \cite{Nogues} in his article
about the great problems of statistical physics for this century.

With our numbers $Y_n$, the produced random walks possess the correct
properties, including the mean distance behavior $<d^2> \sim N$ (see Fig. 4).

\section{Random maps}

The functions of (\ref{Eq2}) are important not only for numerical simulations.
They are relevant by themselves as theoretical paradigms of stochastic
processes \cite{James}. Considering the fact that these are explicit
functions, we can use them to solve (analytically) many theoretical problems
in stochastic dynamical systems.

Consider the following random map \cite{Paladin}:

\begin{equation}
X_{n+1}=\frac{1}{2}\left[ 1+I_{n}\left( 1-4X_{n}\right) \left(
1-X_{n}\right) ^{1/2}\right],  \label{Eq6}
\end{equation}
where $I_{n}$ is a random variable that takes the values $\pm 1$ with equal
probability.

An exact solution to this random map can be written as follows:

\begin{equation}
X_{n}=\sin ^{2}\left[ \theta \pi \left( 3/2\right) ^{n}\right].  \label{Eq7}
\end{equation}

Now we can check some of the results discussed by the authors of Refs.
6 and 7. They have introduced a measure of complexity $K$ in terms
of the average number of bits per time unit necessary to specify the
sequence generated by the system. In dynamical systems of type (\ref{Eq1})
[Eq. (\ref{Eq6} is an example] this measure coincides with the rate $K$ of
divergence of nearby trajectories evolving under two different realizations
of the random variable $I_{n}$.

The complexity of the dynamics can be measured as

\begin{equation}
K=\lambda \theta \left( \lambda \right) +h,  \label{Eq8}
\end{equation}
where $\lambda $ is the Lyapunov exponent of the map, $h$ is the complexity
of $I_{n}$, and $\theta (\lambda )$ is the Heaviside step function.
Complexity $h$ should be defined also as the average number of bits per time
unit necessary to specify the random variable $I_{n}$. When $I_{n}$ is a
usual chaotic noise, then $h$ coincides with the Kolmogorov-Sinai entropy.

In the case of the random map (\ref{Eq6}) $\lambda =\ln \left( 3/2\right) $
and $h=\ln 2$. Hence, $K=\ln 3$. On the other hand, any calculation
(theoretical or numerical) of $K$ for the dynamics generated by function (%
\ref{Eq7}) yields the correct value $K=\ln 3$. Moreover, even an independent
calculation of the complexity of this dynamics using different methods \cite
{Eckmann,Pincus} produces the same result.

Using the functions (\ref{Eq2}) we can also solve the map

\begin{equation}
X_{n+1}=\frac{1+I_{n}\sqrt{1-X_{n}}}{2},  \label{Eq9}
\end{equation}
where $I_{n}$ is defined as in Eq. (\ref{Eq6}).

Here the Lyapunov exponent is negative: $\lambda =\ln \left( 1/2\right) <0$.
However, the complexity is positive: $K=\ln 2$.

In the presence of random perturbations, $K$ can be very different from the
standard Lyapunov exponent and, hence, from the Kolmogorov entropy computed
with the same realization of the randomness.

We stress that a negative value of $\lambda $ does not imply predictability.

In general, if we apply the measure of complexity $K$ to our function (\ref
{Eq2}), then we obtain the following results: For $z=p/q$ ($p$ and $q$
relative primes), $K=\ln p$. If $z$ is irrational, the complexity is
infinite!.

We should say that using function (\ref{Eq2}) we can create complete sets
of orthogonal elements. In the same way that we can solve, to begin with,
any map of type $X_{n+1}=f\left( X_{n}\right) $, we can also solve
theoretically many important problems in stochastic dynamical systems.

\section{Nonlinear forecasting methods}

Now we address the problem of deciding which of the proposed methods \cite
{Sugihara,Wales,Tsonis} for distinguishing chaos from random time series are
more effective. Recently a new method based on nonlinear forecasting was
proposed \cite{Sugihara,Wales,Tsonis}. The idea of the method is as follows.
One can make short-term predictions that are based on a library of past
patterns in a time series (the method of nonlinear forecasting is described
in Refs. 8-10 and the references quoted therein). By
comparing the predicted and actual values, one can make distinctions between
random sequences and deterministic chaos.

For chaotic (but correlated) time series, the accuracy of the nonlinear
forecast falls off with increasing prediction-time interval. On the other
hand, for truly random sequences, the forecasting accuracy is independent of
the prediction interval. The decrease with time of the correlation
coefficient between predicted and actual values can be used to calculate the
largest positive Lyapunov exponent of the time series \cite{Wales}.

Function (\ref{Eq2}) is a very good model system to check this and
other methods. In fact, for $z$ integer, these are chaotic sequences of
type $X_{n+1}=f\left( X_{n}\right) $. For $z=m^{1/k}$, we have chaotic maps
of type: $X_{n+1}=g\left( X_{n},X_{n-1},...,X_{n-k+1}\right) $. For $z$
fractionary, we have different types of random sequences with different
complexities. Finally, for $z$ irrational (generic), the sequence is
maximally random.

Suppose we have a sequence $U_{1},U_{2},...,U_{N}$. Now we construct a map
with the dependence $U_{n}^{predicted}$ as a ``function'' of $%
U_{n}^{observed}$.

If we have a correlated chaotic sequence, this dependence is a straight
line, i.e., $U_{n}^{predicted}\approx $ $U_{n}^{observed}$ (when the
forecasting method is applied for one time step into the future). When we
increase the number of time steps into the future, this relation
deteriorates. When we apply this method to function (\ref{Eq2}) with $z=\pi $
[after transformation $Y_{n}=(2/\pi )\arcsin \sqrt{X_{n}}$], even the map $%
U_{n}^{predicted}$ vs $U_{n}^{observed}$ for one time step into the future
is a map equivalent to that shown in Fig. 5. On the other hand, the
correlation coefficient is independent of the prediction time. In fact,
there are no correlations. The details will be given elsewhere.
Nevertheless, we should say that this method is quite efficient in
distinguishing chaos from randomness. However, it cannot distinguish between
different random time-series.

Other methods discussed in Ref. 8, are less effective in this task.
They are more qualitative, requiring subjective judgment about whether
there is an attractor of given dimensions.

\section{Stochastic resonance}

\subsection{Lyapunov exponents}

A phenomenon that has awakened very much interest in last several years is
stochastic resonance (SR) \cite
{Benzi,McNamara,Bulsara,Gammaitoni,Wiesenfeld,Ippen,Carrol,Kapitaniak,Yang,Crisanti,Anishchenko,Hanggi,Gammaitoni2,Gingl,Chapeau-Blondeau,Bezrukov,Vilar,Marchesoni,González3}%
. The classical model for this phenomenon is the following:

\begin{equation}
\dot{x}-x+x^{3}=A_{0}\sin \left( \omega t\right) +\eta \left( t\right) .
\label{Eq10}
\end{equation}

The sum of a noise signal, $\eta \left( t\right) $, and a weak periodic
signal is used to drive a bistable system. The most important characteristic
of SR is that the signal-to-noise ratio (SNR) has a maximum in the plot SNR
vs $D$, where $D$ is the noise intensity, for a finite nonzero value of the
noise intensity.

It has been shown that SR still occurs when chaos, rather than noise, is
used as the nonperiodic component of the driving signal \cite{Ippen}.
Several authors have investigated the SR in chaotic systems \cite
{Ippen,Carrol,Kapitaniak,Yang,Crisanti,Anishchenko}.

A very important question is how SR depends on the largest Lyapunov exponent
of the driving chaotic noise \cite{Gammaitoni}. We address this problem
systematically for the first time, since we can solve exactly the problem of
calculating the Lyapunov exponent. We should say that for large values of
the Lyapunov exponent, the SR is not a very sensitive phenomenon on the
level of chaos (this is unlike the phenomena discussed in Refs. 1-3.
The curve SNR vs $D$ practically has no
variation for $\lambda \gg \ln 3$. However, in the interval $\ln \left(
3/2\right) <\lambda <\ln 2$, the SR strongly depends on the Lyapunov
exponent. In this interval, the maximum of SNR is shifted to the right
(larger noise intensities) and is amplified!. Figure 6 shows the dependence 
SNR$(D)$ for different values of the Lyapunov exponent $\lambda $. These data
are the result of numerical simulations of Eq. (\ref{Eq10}). We can
compare this result with that obtained in Ref. 33. In this work,
the phenomenon of stochastic resonance is studied in the presence of colored
noise. In overdamped systems, the authors find that SR is suppressed with
increasing noise color. In contrast, for colored noise induced by inertia
(as well as for asymmetric dichotomic noise), they obtain an enhancement of
SR.

The same result can be obtained in the so-called threshold systems \cite
{Gammaitoni2,Gingl,Chapeau-Blondeau}. For instance, define

\begin{equation}
I_{n}=g\left( p_{n}+\eta _{n}\right) ,  \label{Eq11}
\end{equation}
where $p_{n}$ is a periodic function, $\eta _{n}$ is some kind of noise, and 
$g\left( x\right) $ is a function with some properties that allow the
existence of SR \cite{Gammaitoni2,Gingl,Chapeau-Blondeau}.

The simplest case is the following:

\begin{eqnarray}
g\left( x\right) =\cases{-V, x<x_{th}, \cr V, x>x_{th}. }  \label{Eq12}
\end{eqnarray}

A nonlinear circuit with this kind of threshold nonlinearity is discussed in 
Ref. 36.

Different measures have been used to characterize stochastic resonance. In
particular, in Ref. 41 the dynamics of noisy bistable systems is
analyzed by means of Lyapunov exponents and measures of complexity.

It can be shown that, in stochastic resonance systems, the function $K\left(
D\right) $ (where $K$ is the complexity as defined previously and $D$ is the
noise intensity) has a local minimum for a nonzero value of $D$. This
minimum represents an optimal value of noise intensity at which SR occurs.
This result confirms the findings of Ref. 41.

We can consider $K$ as a ``dynamical measure'' of SR because $K$ coincides
with the rate of divergence of nearby trajectories evolving under two
different noise realizations.

This measure can be used to characterize more general stochastic dynamical
systems such as the following:

\begin{equation}
X_{n+1}=F(X_{n},p_{n},\eta _{n}),  \label{EqA}
\end{equation}
where $p_{n}$ is a periodic function and $\eta _{n}$ is the noise.

Let us investigate a particular example:

\begin{equation}
X_{n+1}=\cos \left\{ \left[ 1+\varepsilon \left( p_{n}+D\eta _{n}\right)
^{2}\right] \arccos \left( X_{n}\right) \right\} .  \label{EqB}
\end{equation}

Here $p_{n}$ is a periodic function of amplitude $\alpha $ and $\eta _{n}$
is a chaotic noise defined as follows:

\begin{equation}
\eta _{n}=Y_{n}-\delta -1,  \label{EqC}
\end{equation}
where $\delta $ is a parameter for which $0<\delta <1$ and $Y_{n+1}=\sin
^{2}\left( z\arcsin \sqrt{Y_{n}}\right) $.

For this dynamical system, function $K\left( D\right) $ has a minimum for a
finite $D$.

Let us suppose that $p_{n}$ is a period-one function. We can write down an
analytic expression for $K$:

\begin{equation}
K=\ln \left\{ z\left[ 1+\frac{\varepsilon }{2}\left( \left( \alpha -D\delta
\right) ^{2}+\left( \alpha -D\left( z-1\right) \right) ^{2}\right) \right]
\right\} .  \label{EqD}
\end{equation}

The function $K\left( D\right) $ is shown in Fig. 7. For a fixed $z$, the
minimum of $K$ is obtained approximately for $D=\alpha /\left( z-1\right) $.
If we minimize $K$ with respect to both $D$ and $z$, we obtain $D=\alpha
/\delta $, $z=1+\delta $.

Note that for $1+\delta <z<3$, the minimum is deeper and is shifted to the
right as $z$ is decreased. This is a phenomenon similar to that obtained
using SNR in Fig. 6. See also subsection VI E, where some experiments are
mentioned.

We should say that $K$ can characterize the dynamics of dynamical systems of
type (\ref{EqA}) even when there is no periodic function at all.

In some cases, for a finite value of $D$ we can find the least complex
dynamics.

In some sense, this is a more general phenomenon than the usual stochastic
resonance. In fact, this is an example of the so-called noise-induced
disorder-order transitions, of which the SR phenomena can be a subset.

In Fig. 8, different time series and return maps produced by the dynamical
system (\ref{EqB}) are shown. Note that for some intermediate value of $D$
we obtain the least complex dynamics.

In Fig. 9 we see that when both $D$ and $z$ are very close to the optimal
values, then the resulting dynamics is very predictable.

Note that this system can be chaotic even when $D=0$, due to the intrinsic
nonlinear dynamics of the system. However, for some finite value of the
noise intensity $D>0$, we can control this chaotic dynamics. So, in this
case, we are truly controlling the chaotic system using chaotic noise.

\subsection{Explicit output functions for SR systems}

Recently scientists have learned that stochastic resonance can appear not
only in bistable systems \cite{Chapeau-Blondeau,Bezrukov,Vilar}. A very
interesting class of systems is that of the so-called nonlinear static (or
``nondynamical'') systems. In Refs. 36-38 a
theory of these systems is presented. Using this theory and our function (%
\ref{Eq2}) we can write down an explicit ``solution'' function to these
systems. For instance, the function:

\begin{equation}
I_{n}=\tanh \left\{ B\left[ A_{0}\sin \left( \omega n\right) +D\cos \left(
\theta \pi z^{n}\right) -V_{th}\right] \right\}  \label{Eq13}
\end{equation}
can behave as a SR system. Figure 10 shows that the function (\ref{Eq13}) is
a SR system for $B=24$. For $B=1$ the stochastic resonance disappears. Here
the SNR was calculated numerically using function (\ref{Eq13}) as the output
signal. In fact, we can construct a very general class of SR functions of
type $I_{n}=g\left( V_{n}\right) +\xi _{n}$, where $V_{n}=p_{n}+\eta _{n}$
is the input and $I_{n}$ is the output. Function $\xi _{n}$ represents the
intrinsic noise \cite{Vilar}.

Note that although the systems described in Refs. 36-38
are called ``static systems,'' once we
have constructed our explicit functions, e.g., Eq. (\ref{Eq13}), we can
obtain exact solutions to very dynamical systems. Also note that it is very
easy to check that there is a maximum in the dependence SNR vs $D$; however
our explicit functions can be investigated using mathematical analysis, not
only statistics. Very different functions with very different dynamics can
have the same SNR behavior and other statistical properties. Using our
explicit functions we can investigate the true dynamics of the system.

The analysis of function (\ref{Eq2}) allows us to construct a continuous
and differentiable function with properties similar to those of the chaotic
functions (\ref{Eq2}). Let us give an example:

\begin{equation}
\begin{array}{c}
f\left( t\right) =\sin \{B_{1}\sinh \left[ a_{1}\cos \left( \omega
_{1}t\right) +a_{2}\cos \left( \omega _{2}t\right) \right] + \\ 
B_{2}\cosh \left[ a_{3}\cos \left( \omega _{3}t\right) +a_{4}\cos \left(
\omega _{4}t\right) \right] \}.
\end{array}
\label{Eq14}
\end{equation}

Using functions of this kind we can find analytic solutions to continuous
chaotic dynamical systems. Function (\ref{Eq14}) with the parameter values $%
B_{1}=20,B_{2}=30,a_{1}=10,a_{2}=15,a_{3}=10,a_{4}=15,\omega _{1}=1,\omega
_{2}=\pi ,\omega _{3}=\sqrt{2},\omega _{4}=e$; behaves as a chaotic system
(see Fig. 11). Any investigation (theoretical or numerical) will give the
same result: The maximum Lyapunov exponent is positive. Moreover, if we need
a continuous dynamics with a chaotic Gaussian-like ``noise'' we can use a
transformation of Eq. (\ref{Eq14}): $g(t)=\ln \left[ f^{2}(t)/\left(
1-f^{2}(t)\right) \right] $. We have been able to produce SR with function (%
\ref{Eq14}), $f(t)$, and $g(t)$. In Fig. 12 SNR is calculated from numerical
simulations of Eq. (\ref{Eq10}) and using the continuous chaotic
function $g(t)=\ln \left[ f^{2}(t)/\left( 1-f^{2}(t)\right) \right] $, where 
$f(t)$ is defined by Eq. (\ref{Eq14}).

\subsection{Solitonic stochastic resonance}

The spatiotemporal SR in the $\varphi ^{4}$ model has been considered in a
very interesting paper \cite{Marchesoni}. Recently we introduced the concept
of solitonic stochastic resonance (SSR) \cite{González3} where a soliton
moves in a bistable potential created by space-dependent external forces
driven by a periodic signal and noise. This seems to be equivalent to the
conventional setup for SR, however the conditions for the existence of SSR
are different from that of SR with a point particle in a bistable potential.
The situation for SSR can produce very interesting phenomena like the
transformation of the soliton into a three-``particle'' system of two
solitons and an antisoliton \cite{González3}.

Here we will present another framework for SSR.

The function

\begin{equation}
\phi \left( x,t\right) =\tanh \left\{ B\left[ x-x_{0}-A\sin \left( \omega
t\right) -Df\left( t\right) \right] \right\} ,  \label{Eq15}
\end{equation}
where $f(t)$ is defined in Eq. (\ref{Eq14}), can be used to find analytic
solutions to nonlinear partial differential equations.

Recall that function (\ref{Eq15}) is a SR solution [see Eq. (\ref{Eq13})].
For instance, if we take the time series produced by $\phi \left(
x=0,t\right) $ [with $B=12,A=0.67,\omega =0.88,x_{0}=2,$ and $f(t)$ is the
function (\ref{Eq14})] we obtain a new kind of SSR from (\ref{Eq15}). In
fact, using the solution (\ref{Eq15}) we can prove that the overdamped
perturbed $\varphi ^{4}$ equation

\begin{equation}
\phi _{t}-\phi _{xx}-B^{2}\left( \phi -\phi ^{3}\right) =\frac{B\left[
\omega A\cos \left( \omega t\right) +D\dot{f}(t)\right] }{\cosh ^{2}\left[
B\left( x-x_{0}\right) \right] }+F(x),  \label{Eq17}
\end{equation}
where $F(x)=a\tanh \left[ B\left( x-x_{0}\right) \right] $, possesses a
different kind of SSR.

We can calculate analytically the SNR for the dynamics of Eq. (\ref
{Eq17}).

Suppose that, in Eq. (\ref{Eq17}), instead of $f(t)$, the noise is
described by the function $h(t)=(2/\pi)\arcsin f(t)$, and $f(t)$ is given by
function (\ref{Eq14}). This is equivalent to a uniformly distributed noise
in the interval $(0,1)$. We will define $b/2=x_0-A$.

Following the ideas of Ref. 34 we can calculate an
approximate analytical expression for the SNR when $B\gg 1$: 
\[
SNR=\cases{0 \hspace{2mm} for \hspace{2mm} \frac{D}{2}<\frac{b}{2}-A \cr
\frac{1}{D}(\frac{D}{2}-\frac{1}{2}+2A) \hspace{2mm} for \hspace{2mm} \frac{b%
}{2}-A\le\frac{D}{2}\le \frac{b}{2}+A\cr \frac{2A}{D} \hspace{2mm} for 
\hspace{2mm} \frac{D}{2}> \frac{b}{2}+A } 
\]

It is evident that there is a maximum in the curve SNR$(D)$.

Our theoretical results on the theory of solitons perturbed by external
forces \cite{González4,González5,González6,González7} allow us to understand
the dynamics of Eq. (\ref{Eq17}) and to interpret the physics of solution (%
\ref{Eq15}). In order to obtain the desired dynamics we should solve
Eq. (\ref{Eq17}) with an initial condition representing a soliton
situated in a vicinity of point $x=x_{0}$. In this case, the soliton center
of mass will be oscillating inside the potential well created by the force $%
F(x)$. This is exactly what represents solution (\ref{Eq15}).

When we investigate the time series obtained after the numerical simulation
of Eq. (\ref{Eq17}), we obtain SSR as predicted by the theoretical solution (%
\ref{Eq15}).

The SNR vs $D$ plot depends on the value of $B$. For very large values of $B$%
, the SNR vs $D$ plot has a very nice maximum (see Fig. 13). The SNR$(D)$
dependence shown in Fig. 13 was calculated numerically from the time series
generated by the function $\phi (x=0,t)$ as a solution of Eq. (\ref{Eq17}).
The same result is obtained if we investigate the analytic solution (\ref
{Eq15}). For very small values of $B$, the SSR disappears. Thus, this is a
SSR that depends on the shape of the soliton. In particular, it depends on
the width of the soliton, which can be expressed as $S=1/B$. This SSR is
different from the one obtained for a soliton moving in a bistable potential
well \cite{González3} and that described in Ref. 39.

In Ref. 39 the synchronization of a linearly coupled chain of 
$N$ overdamped bistable elements, subject to a deterministic periodic signal
and uncorrelated white noise, is addressed in the continuous limit of a $%
\varphi ^{4}$ theory. The cooperation between noise and coupling is shown to
lead to spatiotemporal stochastic resonance. There, the bistability of the $%
\varphi ^{4}$ equation on the potential $U(\varphi )\sim (\varphi
^{2}-1)^{2} $ plays the most important role. On the other hand, in our
previous paper \cite{González3} we considered the stochastic resonance of a
soliton moving in a bistable potential created by inhomogeneous external
forces $F(x)$. In this paper, the output signal is the coordinate of the
soliton center of mass. In the present work, the relevant output signal is $%
\phi (x=0,t)$. The soliton is moving in a monostable well potential created
by inhomogeneous external forces. However, the most striking feature is that
the width of the soliton determines the existence or not of the solitonic
stochastic resonance. The shape of the output signal possesses patterns that
are very different from that obtained in a bistable system. They are more
similar to the patterns that appear in threshold systems.

\subsection{Patterns in the output signal of SR systems}

Once we have an explicit solution that describes the stochastic resonance
system as the following: 
\begin{equation}
I(t)=g\left( P(t)+\eta (t)\right)
\end{equation}
and 
\begin{equation}
I(t)=g\left( P(t)+\eta (t)\right) +\xi (t),
\end{equation}
where $P(t)$ is a periodic function, $\eta (t)$ and $\xi (t)$ are different
manifestations of noise dynamics, we can calculate the SNR exactly. We
should note that SNR is the main measure of stochastic resonance and is
widely used in SR literature.

For instance, let us define the different noises as follows. $\xi(t)$ is a
white noise with zero mean and correlation function: 
\begin{equation}
<\xi(t)\xi(t+\tau)>=Q\delta(\tau),
\end{equation}
where Q is a constant parameter. $\eta(t)$ is a Gaussian noise with zero
mean and correlation function: 
\begin{equation}
<\eta(t)\eta(t+\tau)>=\sigma^2 exp(-\tau/\tau_F).
\end{equation}

Following Ref. 38 we can consider the case 
\begin{equation}
g(V)=V^{3},
\end{equation}
where 
\begin{equation}
V=\alpha \sin (w_{0}t)+\eta (t).
\end{equation}

In this case the SNR is 
\begin{equation}
SNR=\pi\frac{18\alpha^2\sigma^4+9\alpha^4\sigma^2+(9/8)\alpha^6} {%
2Q+\tau_F\left( 44\sigma^6+54\alpha^2\sigma^4+(27/2)\alpha^4\sigma^2 \right)}%
.
\end{equation}

This is exactly the SNR obtained in Ref. 38.

The static character of the present nonlinearities allows a direct
statistical analysis, in which all quantities relevant to characterize the
SNR in the output signal can be obtained from statistics computed directly
on the input noises. In fact the SNR is a statistical measure based on the
statistical properties of noises $\eta(t)$ and $\xi(t)$.

Nevertheless, we believe that using our explicit functions we can obtain
much more information about the output signal. Some of this information can
have statistical character, but we will have also dynamical and geometrical
information about the output signal.

For instance, we can predict the values of the local maxima and minima in
the time series, and the distance between them. We can obtain the exact
analytical shape of the extrema.

In any stochastic resonance output signal there are patterns. These patterns
can be different for different systems. The following function 
\begin{equation}
I_{n}=\frac{tanh[B(A\sin (wn)+D\cos (\theta \pi z^{n}))-V_{th}]+1}{2}
\label{InSR}
\end{equation}
is the analytical solution for a circuit with $I-V$ characteristic of type: 
\[
g(V)=\cases{0\hspace{2mm}for\hspace{2mm}V<V_{th}\cr 1\hspace{2mm}for\hspace{%
2mm}V>V_{th}}
\]
(see Ref. 36). A typical time series is shown in Fig.
14(a).

In Fig. 14(b) is shown the typical time series for
\begin{equation}
g(V)=\cases{-1 \hspace{2mm} for \hspace{2mm} V<-0.5 \cr 0 \hspace{2mm} for
\hspace{2mm} -0.5\le V \le 0.5 \cr 1 \hspace{2mm} for \hspace{2mm} V>0.5 }
\label{eq333}
\end{equation}
In Fig. 14(c) we show the function 
\begin{equation}
I_{n}=\left( \frac{tanh(BV_{n})+1}{2}\right) bV_{n},  \label{InRE2}
\end{equation}
where 
\begin{equation}
V_{n}=A\sin (wn)+D\cos (\theta \pi z^{n}),
\end{equation}
which is the output signal for a circuit system with the characteristic
\begin{equation}
g(V)=\cases{0 \hspace{2mm} for \hspace{2mm} V<V_{th}\cr
b(V-V_{th})\hspace{2mm} for \hspace{2mm} V>V_{th} }.
\end{equation}
Figure 14(d) shows the output signal 
\begin{equation}
I_{n}=g(V_{n})+\xi _{n},  \label{Seq}
\end{equation}
where $g(V)=V^{3}$, $V_{n}=\alpha \sin (wn)+\eta _{n}$, $\eta _{n}=DY_{n}$, $%
Y_{n}=ln\left( \frac{X_{n}}{1-X_{n}}\right) $, $X_{n}=\sin ^{2}\left( \theta
\pi z^{n}\right) $, $\xi _{n}=Qln\frac{Z_{n}}{1-Z_{n}}$, and $Z_{n}=\sin
^{2}\left( \theta \pi \pi ^{n}\right) $.

All these systems present stochastic resonance. All these systems can be
tuned to have the same SNR. However, note that all the patterns are
different. Compare them to the typical time series of a classic bistable
stochastic resonance system shown in Fig. 15. The information about these
patterns is in the explicit functions that can be written down using our
stochastic functions.

We can even make predictions about the outcomes in these stochastic systems.
For instance, in the function (\ref{InSR}) we can say that with a
probability $p=0.8$, after the function $I_n$ has taken the value $I_n=1$,
it will take the value $I_n=0$. Meanwhile, we can expect that it will remain
in the state $I_n=0$ for an average time aproximately equal to the period of
the periodic input signal.

On the other hand, the output function (\ref{InRE2}) will give us much more
information about the actual shape of the input periodic signal than the
functions (\ref{InSR}) and (\ref{eq333}).

Systems with intrinsic and external noises are expected to be very random.
Nevertheless, using the theoretical information obtained from our explicit
function (\ref{Seq}), we can make very remarkable predictions. For instance,
if the output signal takes a ``large'' negative value (say $I_{n}=-20$),
then with absolute certainty we can predict that the next values will be
negative and $|I_{n}|\to 0$. When $I_{n}$ reaches the value $I_{n}=0$, we can
predict that the next value will be positive, but the exact value is
unpredictable. When it takes this positive value, the next value will be
negative with absolute certainty. The larger the absolute value of $I_{n}$
when it takes a positive value, the larger the absolute value of the next
negative value that it will take. Note that all the randomness of this
dynamics is produced when $I_{n}$ is near zero. When $I_{n}$ is far from
zero, we can make exact predictions of the next values. All this can be
corroborated when we observe the first-return map of this dynamics (Fig. 16).

We can see the stochastic resonance as a phenomenon that transforms a
complex dynamics into a simpler one. That is, the output signal is less
complex that the input signal. But there are also phenomena that lead to a
more complex behavior (e.g., the chaotic systems).

Using our functions we can predict the existence of new complex phenomena.

After an analysis of the functions $X_n=\sin^2\left(\theta\pi z^n\right)$,
which we have shown to produce complex dynamics, the first characterics that
surface are the following: The function can be rewritten in the form $%
X_n=h(f(n))$, where the argument function $f(n)$ grows exponentially and the
function $h(y)$ is always finite and periodic.

However, a more thorough analysis shows that (to produce complex behavior) the
function $f(n)$ does not have to be exponential all the time, and the function $h(y)$
does not have to be periodic.

In fact, it is sufficient that the function $f(n)$ be a nonperiodic
oscillating function where there are repeating intervals with finite
exponetial behavior. For instance, this can be a chaotic function. On the
other hand, function $h(y)$ should be noninvertible. In other words, it
should have different maxima and minima. The inverse ``function'' of $h(y)$
should be multivalued.

The complexity of the output dynamics is proportional to the number of
extrema of function $h(y)$.

For example, the following system can produce a dynamics similar to that
obtained with our function $X_{n}=\sin ^{2}\left( \theta \pi z^{n}\right) $: 
\begin{eqnarray}
X_{n+1}= \cases{aX_n &if $X_n<Q$\cr bY_n &if $X_n>Q$,}  \label{eq:probl}
\end{eqnarray}
\begin{equation}
Y_{n+1}=\sin ^{2}(d\arcsin \sqrt{Y_{n}}), \label{eq:probl2}
\end{equation}
\begin{equation}
Z_{n+1}=g(X_{n}), \label{eq:probl3}
\end{equation}   
where $g(X_{n})$ is a function with several maxima and minima. The first
return map is shown in Fig. 17. In fact, this is a completely new chaotic
phenomenon because the dynamics is completely unpredictable. So, when the
input is a simple chaotic signal and the system is an electronic circuit
with the $I-V$ characteristic shown in Fig. 18, then we will have a very
complex output. This phenomenon is the opposite to the stochastic resonance.
Compare two $I-V$ characteristic curves for a phenomenon that simplifies the
dynamics and for a new phenomenon that makes the dynamics extremely complex
in Fig. 18. In Ref. 46 a theory of nonlinear circuits is presented.
There we can find different methods to construct circuits with these $I-V$
characteristic curves.

All the results presented in Sec. VI D, which are related to the
Figs. 14-18 were obtained through theoretical calculations.

\subsection{Applications in real systems}

In this section we will present some examples that show how our technique
can be used in real world applications.

Our group have designed and constructed a nonlinear circuit (using a concave
resistor) with the $I-V$ characteristic described by Eq. (36) (see Ref. 
46). We wished to check our theoretical results about the
dependence of SNR on the Lyapunov exponent of the chaotic noise. We also
desired to observe the patterns for the output signal predicted by our
theory.

In order to have different driving chaotic signals, we produced numerical
time series using the exactly solvable map (\ref{Eq3}) for different $z$.

Then, we transformed the numerical time series into analog signals using a
converter. These analog signals plus a subthreshold periodic signal were
introduced as the voltage to the concave resistor circuit. The current in
the concave resistor was taken as the output signal.

We should say that the amplification of the SNR in the interval $\ln \left(
3/2\right) <\lambda <\ln 2$ was clearly observed. The maximum SNR for $%
\lambda =\ln \left( 3/2\right) $ was 5 times larger than for $\lambda \geq 3$
or for any random noise.

We will present further details of these experiments elsewhere.

Thus, in the case that SR is used for the amplification of small signals (as
in the dithering effect) such that the added external noise is a
controllable and manipulable parameter, our recommendation is to utilize a
uniform chaotic noise with a Lyapunov exponent of the order of $\lambda
\approx \ln \left( 3/2\right) $.

Using the techniques described in Ref. 46 it is possible to
construct nonlinear circuits with the $I-V$ characteristics shown in Fig. 18.

For our experiments we used the twin-transistor circuit \cite{Chua}.

As an input signal (voltage), we introduced an analog chaotic signal
previously produced by a nonlinear map.

Playing with different parameters we were able to produce different
unpredictable dynamics very similar to those obtained from our function (\ref
{Eq2}) and the Figs. 2 and 3. The same results can be obtained when we take
the input signal from a chaotic electronic circuit.

In general, the patterns or absence of patterns (it depends on the
nonlinearity) predicted by our theory are completely confirmed by the
experiments.

Some of these experiments are complicated and should be explained in a
separate paper.

In many relevant applications it is important to get resonances or (in other
cases) to avoid resonances.

Recently a new resonance concept was introduced: the geometrical resonance%
\cite{Chacón,González3}.

In the usual linear resonance phenomena, the amplitude and frequency of the
driving force are the most important characteristics. However, in nonlinear
systems the shape of the driving signal becomes crucial. The geometrical
resonance considers the amplitude, the frequency, and the shape of the
perturbation.

Suppose we have a nonlinear system $A$ which for some application should be
driven by a specific driving signal with a given shape. In that case, we can
use the output signal of another system $B$ as the input signal of the
system $A$. So it is important to predict the shape of the output of
nonlinear systems. With such information we can design the appropriate system
to produce the desired signal needed for the driving of the system $A$.

In Sec. VI D we have shown that we can predict specific patterns and
the shape of the output signal of nonlinear stochastic systems. This is a
step forward in the control of chaotic and stochastic systems.

In general, it is very important to find patterns and regularities in the
stochastic dynamical systems. In fact, not everything that can be observed
can be predicted, only the regularities in the observations are the
``province of science'' \cite{Hartle}.

There are many stochastic systems (including systems presenting stochastic
resonance) where predictions are crucial. In this case we mean
predictions of the true values of the outcomes using the previous values.
Among the concerned areas are the following: geophysics, meteorology,
climatology, social sciences, etc. \cite{Stone,Saltzman,Liu,Babinec}

In Sec. VI D we have investigated a system with external noise and
intrinsic noise. These noisy perturbations are very unpredictable functions.
However, we have shown theoretically that we are able to make direct
predictions of the time-series outcomes. There are many systems with this
behavior \cite{Vilar}. When we observe Fig. 14(d), we note that there are
very remarkable bursts in the time series. In all the mentioned applications
it is very important to predict these bursts. We have shown that we can do
this.

\section{Conclusion}

In conclusion, we can construct functions that are exact solutions to
chaotic dynamical systems. Moreover, we have generalized functions that
cannot be generated by a finite recursive algorithm. They can be utilized as
theoretical paradigms of stochastic processes. Considering the fact that
these are explicit functions, we can use them to solve (analytically) many
theoretical problems in stochastic dynamical systems. Thus, we can apply
dynamical concepts to describe processes that are usually considered only
statistically\cite{Carrol}.

\section{Acknowledgments}

Jos\'{e} Juan Su\'{a}rez (UNEXPO, Venezuela) and Dr. Gustavo
Guti\'{e}rrez (USB, Venezuela) have collaborated in the design and
construction of the nonlinear circuits.

\newpage

\begin{table}[tbp]
\caption{Representation of the matrix $X_{n}^{k}$ given by Eq. (5) with $z=2$
and $\theta_{0}=2^{1/2}-1$. Note that if we start with the same initial
conditions, then we will have the same chaotic sequences.}{\scriptsize %
\tabcolsep 0.06cm 
\begin{tabular}{c||c|c|c|c|c|c|c|c|c|c|c|c|c|c|c}
& $\theta_0$ & $\theta_0+1$ & $\theta_0+2$ & $\theta_0+3$ & $\theta_0+4$ & $%
\theta_0+5$ & $\theta_0+6$ & $\theta_0+7$ & $\theta_0+8$ & $\theta_0+9$ & $%
\theta_0+10$ & $\theta_0+11$ & $\theta_0+12$ &  &  \\ \hline\hline
$X_0$ & 0.9291 & 0.9291 & 0.9291 & 0.9291 & 0.9291 & 0.9291 & 0.9291 & 0.9291
& 0.9291 & 0.9291 & 0.9291 & 0.9291 & 0.9291 &  &  \\ 
$X_1$ & 0.2634 & 0.2634 & 0.2634 & 0.2634 & 0.2634 & 0.2634 & 0.2634 & 0.2634
& 0.2634 & 0.2634 & 0.2634 & 0.2634 & 0.2634 &  &  \\ 
$X_2$ & 0.7762 & 0.7762 & 0.7762 & 0.7762 & 0.7762 & 0.7762 & 0.7762 & 0.7762
& 0.7762 & 0.7762 & 0.7762 & 0.7762 & 0.7762 &  &  \\ 
$X_3$ & 0.6948 & 0.6948 & 0.6948 & 0.6948 & 0.6948 & 0.6948 & 0.6948 & 0.6948
& 0.6948 & 0.6948 & 0.6948 & 0.6948 & 0.6948 &  &  \\ 
$X_4$ & 0.8481 & 0.8481 & 0.8481 & 0.8481 & 0.8481 & 0.8481 & 0.8481 & 0.8481
& 0.8481 & 0.8481 & 0.8481 & 0.8481 & 0.8481 &  &  \\ 
$X_5$ & 0.5151 & 0.5151 & 0.5151 & 0.5151 & 0.5151 & 0.5151 & 0.5151 & 0.5151
& 0.5151 & 0.5151 & 0.5151 & 0.5151 & 0.5151 &  &  \\ 
$X_6$ & 0.9990 & 0.9990 & 0.9990 & 0.9990 & 0.9990 & 0.9990 & 0.9990 & 0.9990
& 0.9990 & 0.9990 & 0.9990 & 0.9990 & 0.9990 &  &  \\ 
$X_7$ & 0.0036 & 0.0036 & 0.0036 & 0.0036 & 0.0036 & 0.0036 & 0.0036 & 0.0036
& 0.0036 & 0.0036 & 0.0036 & 0.0036 & 0.0036 &  &  \\ 
$X_8$ & 0.0146 & 0.0146 & 0.0146 & 0.0146 & 0.0146 & 0.0146 & 0.0146 & 0.0146
& 0.0146 & 0.0146 & 0.0146 & 0.0146 & 0.0146 &  &  \\ 
$X_9$ & 0.0578 & 0.0578 & 0.0578 & 0.0578 & 0.0578 & 0.0578 & 0.0578 & 0.0578
& 0.0578 & 0.0578 & 0.0578 & 0.0578 & 0.0578 &  &  \\ 
$X_{10}$ & 0.2181 & 0.2181 & 0.2181 & 0.2181 & 0.2181 & 0.2181 & 0.2181 & 
0.2181 & 0.2181 & 0.2181 & 0.2181 & 0.2181 & 0.2181 &  & 
\end{tabular}
}
\end{table}

\begin{table}[tbp]
\caption{Representation of the matrix $X_{n}^{k}$ defined by Eq. (5) with $%
z=3/2$ and $\theta_0=1$. Note that all the column sequences possess the same
initial conditions $X_{0}=0$. However, all the sequences are different in
general.}{\scriptsize \tabcolsep 0.06cm 
\begin{tabular}{c||c|c|c|c|c|c|c|c|c|c|c|c|c|c|c}
& $\theta_0$ & $\theta_0+1$ & $\theta_0+2$ & $\theta_0+3$ & $\theta_0+4$ & $%
\theta_0+5$ & $\theta_0+6$ & $\theta_0+7$ & $\theta_0+8$ & $\theta_0+9$ & $%
\theta_0+10$ & $\theta_0+11$ & $\theta_0+12$ &  &  \\ \hline\hline
$X_0$ & 0 & 0 & 0 & 0 & 0 & 0 & 0 & 0 & 0 & 0 & 0 & 0 & 0 &  &  \\ 
$X_1$ & 1 & 0 & 1 & 0 & 1 & 0 & 1 & 0 & 1 & 0 & 1 & 0 & 1 &  &  \\ 
$X_2$ & 1/2 & 1 & 1/2 & 0 & 1/2 & 1 & 1/2 & 0 & 1/2 & 1 & 1/2 & 0 & 1/2 &  & 
\\ 
$X_3$ & 0.8535 & 1/2 & 0.1464 & 1 & 0.1464 & 1/2 & 0.8535 & 0 & 0.8535 & 1/2
& 0.1464 & 1 & 0.1464 &  &  \\ 
$X_4$ & 0.0380 & 0.1464 & 0.3086 & 1/2 & 0.6913 & 0.8535 & 0.9619 & 1 & 
0.9619 & 0.8535 & 0.6913 & 1/2 & 0.3086 &  &  \\ 
$X_5$ & 0.9157 & 0.3086 & 0.4024 & 0.8535 & 0.0096 & 0.9619 & 0.2222 & 1/2 & 
0.7777 & 0.0380 & 0.9903 & 0.1464 & 0.5975 &  &  \\ 
$X_6$ & 0.8865 & 0.4024 & 0.2643 & 0.9619 & 0.0215 & 0.7777 & 0.5490 & 0.1464
& 0.9975 & 0.0842 & 0.6451 & 0.6913 & 0.0590 &  &  \\ 
$X_7$ & 0.0711 & 0.2643 & 0.5245 & 0.7777 & 0.9519 & 0.9975 & 0.9016 & 0.6913
& 0.4266 & 0.1828 & 0.0292 & 0.0096 & 0.1295 &  &  \\ 
$X_8$ & 0.8447 & 0.5245 & 0.1213 & 0.9975 & 0.1923 & 0.4266 & 0.9087 & 0.0096
& 0.7674 & 0.6214 & 0.0649 & 0.9784 & 0.2751 &  &  \\ 
$X_9$ & 0.9686 & 0.1213 & 0.7410 & 0.4266 & 0.3964 & 0.7674 & 0.1020 & 0.9784
& 0.0009 & 0.9571 & 0.1421 & 0.7137 & 0.4571 &  &  \\ 
$X_{10}$ & 0.7544 & 0.7410 & 0.0002 & 0.7674 & 0.7275 & 0.0009 & 0.7803 & 
0.7137 & 0.0021 & 0.7928 & 0.6998 & 0.0037 & 0.8051 &  & 
\end{tabular}
}
\end{table}

\begin{table}[tbp]
\caption{Representation of the matrix $X_{n}^{k}$ defined by Eq. (5) with $%
z=4/3$ and $\theta_0=1/6$. Note that the horizontal row sequences possess
periods $3^{n}$. All the next-values in the column sequences are
unpredictable.}{\scriptsize \tabcolsep 0.06cm 
\begin{tabular}{c||c|c|c|c|c|c|c|c|c|c|c|c|c|c|c}
& $\theta_0$ & $\theta_0+1$ & $\theta_0+2$ & $\theta_0+3$ & $\theta_0+4$ & $%
\theta_0+5$ & $\theta_0+6$ & $\theta_0+7$ & $\theta_0+8$ & $\theta_0+9$ & $%
\theta_0+10$ & $\theta_0+11$ & $\theta_0+12$ &  &  \\ \hline\hline
$X_0$ & 1/4 & 1/4 & 1/4 & 1/4 & 1/4 & 1/4 & 1/4 & 1/4 & 1/4 & 1/4 & 1/4 & 1/4
& 1/4 &  &  \\ 
$X_1$ & 0.4131 & 0.9698 & 0.1169 & 0.4131 & 0.9698 & 0.1169 & 0.4131 & 0.9698
& 0.1169 & 0.4131 & 0.9698 & 0.1169 & 0.4131 &  &  \\ 
$X_2$ & 0.6434 & 0.0531 & 0.2014 & 0.8431 & 0.9177 & 0.3019 & 0.0134 & 0.5290
& 0.9966 & 0.6434 & 0.0531 & 0.2014 & 0.8431 &  &  \\ 
$X_3$ & 0.8951 & 0.4515 & 0.1712 & 0.9996 & 0.1430 & 0.4903 & 0.8702 & 0.0015
& 0.8139 & 0.5676 & 0.0932 & 0.9906 & 0.2333 &  &  \\ 
$X_4$ & 0.9929 & 0.6920 & 0.2118 & 0.0006 & 0.2556 & 0.7387 & 0.9989 & 0.7933
& 0.3138 & 0.0081 & 0.1616 & 0.6309 & 0.9780 &  &  \\ 
$X_5$ & 0.6475 & 0.0675 & 0.1584 & 0.7792 & 0.9668 & 0.4302 & 0.0018 & 0.3463
& 0.9291 & 0.8462 & 0.2261 & 0.0308 & 0.5633 &  &  \\ 
$X_6$ & 0.0393 & 0.9703 & 0.2695 & 0.3683 & 0.9239 & 0.0086 & 0.7980 & 0.5534
& 0.1235 & 0.9998 & 0.1420 & 0.5262 & 0.8194 &  &  \\ 
$X_7$ & 0.4955 & 0.5309 & 0.4425 & 0.5837 & 0.3901 & 0.6355 & 0.3390 & 0.6858
& 0.2897 & 0.7340 & 0.2427 & 0.7796 & 0.1987 &  &  \\ 
$X_8$ & 0.7550 & 0.7849 & 0.8132 & 0.8400 & 0.8651 & 0.8884 & 0.9097 & 0.9290
& 0.9461 & 0.9609 & 0.9735 & 0.9837 & 0.9914 &  &  \\ 
$X_9$ & 0.4054 & 0.9858 & 0.1903 & 0.2718 & 0.9995 & 0.3124 & 0.1565 & 0.9733
& 0.4495 & 0.0686 & 0.9093 & 0.5907 & 0.0151 &  &  \\ 
$X_{10}$ & 0.0160 & 0.6018 & 0.9938 & 0.4460 & 0.0008 & 0.5054 & 0.9996 & 
0.5430 & 0.0045 & 0.4089 & 0.9866 & 0.6383 & 0.0267 &  & 
\end{tabular}
}
\end{table}

\begin{table}[tbp]
\caption{Representation of the matrix $X_{n}^{k}$ given by Eq. (5) with $%
z=\pi $ and $\theta_0=1/4$. Note that it is difficult even to find
``clusters'' of equal values in different column sequences. All
column sequences are completely random and different.}{\scriptsize %
\tabcolsep 0.06cm 
\begin{tabular}{c||c|c|c|c|c|c|c|c|c|c|c|c|c|c|c}
& $\theta_0$ & $\theta_0+1$ & $\theta_0+2$ & $\theta_0+3$ & $\theta_0+4$ & $%
\theta_0+5$ & $\theta_0+6$ & $\theta_0+7$ & $\theta_0+8$ & $\theta_0+9$ & $%
\theta_0+10$ & $\theta_0+11$ & $\theta_0+12$ &  &  \\ \hline\hline
$X_0$ & 1/2 & 1/2 & 1/2 & 1/2 & 1/2 & 1/2 & 1/2 & 1/2 & 1/2 & 1/2 & 1/2 & 1/2
& 1/2 &  &  \\ 
$X_1$ & 0.3897 & 0.0516 & 0.0457 & 0.3761 & 0.7983 & 0.9995 & 0.8307 & 0.4169
& 0.0646 & 0.0348 & 0.3494 & 0.7756 & 0.9976 &  &  \\ 
$X_2$ & 0.9895 & 0.7599 & 0.3653 & 0.0562 & 0.0286 & 0.3002 & 0.6984 & 0.9708
& 0.9444 & 0.6359 & 0.2412 & 0.0107 & 0.0906 &  &  \\ 
$X_3$ & 0.4950 & 0.4753 & 0.4556 & 0.4360 & 0.4165 & 0.3972 & 0.3780 & 0.3589
& 0.3401 & 0.3216 & 0.3033 & 0.2853 & 0.2677 &  &  \\ 
$X_4$ & 0.7996 & 0.4643 & 0.2603 & 0.9388 & 0.0012 & 0.9002 & 0.3252 & 0.3937
& 0.8535 & 0.0114 & 0.9684 & 0.2003 & 0.5356 &  &  \\ 
$X_5$ & 0.9997 & 0.9940 & 0.9807 & 0.9601 & 0.9324 & 0.8982 & 0.8579 & 0.8120
& 0.7615 & 0.7069 & 0.6492 & 0.5892 & 0.5278 &  &  \\ 
$X_6$ & 0.7869 & 0.5423 & 0.1479 & 0.9978 & 0.0881 & 0.6342 & 0.7058 & 0.0498
& 0.9849 & 0.2060 & 0.4662 & 0.8458 & 0.0030 &  &  \\ 
$X_7$ & 0.0521 & 0.8342 & 0.7685 & 0.0216 & 0.4881 & 0.9847 & 0.2517 & 0.1484
& 0.9368 & 0.6171 & 0.0003 & 0.6509 & 0.9186 &  &  \\ 
$X_8$ & 0.1640 & 0.7578 & 0.3299 & 0.5757 & 0.5214 & 0.3822 & 0.7096 & 0.2064
& 0.8663 & 0.0746 & 0.9681 & 0.0066 & 0.9997 &  &  \\ 
$X_9$ & 0.5777 & 0.8516 & 0.9930 & 0.9485 & 0.7348 & 0.4326 & 0.1558 & 0.0087
& 0.0469 & 0.2559 & 0.5569 & 0.8365 & 0.9891 &  &  \\ 
$X_{10}$ & 0.0013 & 0.0343 & 0.1084 & 0.2171 & 0.3507 & 0.4976 & 0.6446 & 
0.7789 & 0.8885 & 0.9639 & 0.9982 & 0.9885 & 0.9357 &  & 
\end{tabular}
}
\end{table}

\begin{figure}[tbp]
\caption{One-valued first-return map produced by function (\ref{Eq2}) with $%
z=5$.}
\end{figure}

\begin{figure}[tbp]
\caption{Multivalued first-return maps produced by function (\ref{Eq2}): (a) 
$z=3/2$; (b) $z=8/5$.}
\end{figure}

\begin{figure}[tbp]
\caption{Random first-return maps for $z=\pi$: (a) first-return map produced
by function (\ref{Eq2}); (b) first-return map produced by function (\ref{Eq2}%
) and with transformation $Y_{n}=\left( 2/\pi \right) \arcsin \left(
X_{n}^{1/2}\right) $.}
\end{figure}

\begin{figure}[tbp]
\caption{Mean distance vs number of steps for a random walk generated
with the random numbers $Y_{n}=\left( 2/\pi \right) \arcsin \left(
X_{n}^{1/2}\right)$, where $X_n$ is given by function (\ref{Eq2}). }
\end{figure}

\begin{figure}[tbp]
\caption{Predicted values one step into the future vs observed values for
the time series generated by function (\ref{Eq2}) with $z=e$, after the
transformation $Y_{n}=\left( 2/\pi \right) \arcsin \left( X_{n}^{1/2}\right) 
$.}
\end{figure}

\begin{figure}[tbp]
\caption{In the interval $\ln \left( 3/2\right) <\lambda <\ln 2$, the
maximum SNR is shifted to the right and is amplified.}
\end{figure}

\begin{figure}[tbp]
\caption{Function $K\left( D\right) $ as given by Eq. (22). Solid line, 
$z=1.5$; dotted line, $z=2$; dashed line, $z=2.5$; dot-dashed line, $z=3$.
Note that as $z$ is decreased, the minimum of function $K\left( D\right) $
is deeper and is shifted to the right.}
\end{figure}

\begin{figure}[tbp]
\caption{Time-series and first-return maps generated by the noise-driven
dynamical system (20). (a) and (d) $D=0.5$; (b) and (e) $D=2$; (c) and (f) $D=20$.
In all cases $\varepsilon =0.5$, $\alpha =1$, $\delta =0.3$, $z=1.5$.}
\end{figure}

\begin{figure}[tbp]
\caption{ Time series generated by the noise-driven dynamical system (20). (a) 
$D=2$, (b) $D=10$. In all cases $\varepsilon =0.5$, $\alpha =1$, $\delta =0.1$%
, $z=1.1$. Note that in case (b) the control is so good that the output
signal is almost periodic and is confined to a very narrow interval of
values. Note also that in this case the noise intensity is 5 times larger!}
\end{figure}

\begin{figure}[tbp]
\caption{The explicit function (\ref{Eq13}) can behave as a stochastic
resonance system. SNR vs noise intensity ($D$) is shown for $%
B=24$ and $B=1$. Note that for $B=1$ there is no maximum in this plot.}
\end{figure}

\begin{figure}[tbp]
\caption{Chaotic time series generated by the continuous function (\ref{Eq14}%
).}
\end{figure}

\begin{figure}[tbp]
\caption{SNR vs noise intensity ($D$) for the dynamics of system (\ref{Eq10}%
). Here the ``noise'' $\eta (t)$ is defined as $\eta (t)=\ln \left[
f(t)/\left( 1-f(t)\right) \right] $, where $f(t)$ is given by Eq. (\ref{Eq14}).}
\end{figure}

\begin{figure}[tbp]
\caption{SNR vs noise intensity ($D$) for the dynamics of system (\ref{Eq17}%
). The SNR is calculated from the time series generated by the function $%
\phi (x=0,t)$.}
\end{figure}

\begin{figure}[tbp]
\caption{Time series produced by explicit functions that describe different
stochastic resonance systems. (a) function (34), (b) system (35), (c) function
(36), (d) function (39).}
\end{figure}

\begin{figure}[tbp]
\caption{ Typical time series for a bistable stochastic resonance system such as
that described by Eq. (16).}
\end{figure}

\begin{figure}[tbp]
\caption{First-return map produced by function (39), which describes a
stochastic resonance system with intrinsic noise. (a) Situation of stochastic
resonance $D=0.16$. (b) Situation out of stochastic resonance $D=0.3$.}
\end{figure}

\begin{figure}[tbp]
\caption{First-return map produced by the dynamics of variable $Z_n$ in the
dynamical system (40)-(42). (a) Function $g(x)$ possesses 1 local extremum. (b)
Function $g(x)$ possesses 100 local extremas.}
\end{figure}

\begin{figure}[tbp]
\caption{$I-V$ characteristic curves of two nonlinear circuits: (a) With some
apropriately chosen input signal, this circuit can produce a very complex
output signal (see the discussion in the text). (b) If the input signal is
composed of a periodic signal and noise, the output signal will be less
complex than the input signal. In fact, this will be a stochastic resonance
system.}
\end{figure}


\begin{references}
\bibitem{Ferrenberg}  A. M. Ferrenberg, D. P. Landau, and Y. J. Wong, Phys.
Rev. Lett. {\bf 69}, 3382 (1992).

\bibitem{D'Souza}  R. M. D'Souza, Y. Bar-Yam, and M. Kardar, Phys. Rev. E 
{\bf 57}, 5044 (1998).

\bibitem{Nogues}  J. Nogu\'{e}s, J. L. Costa-Kr\"{a}mer, and K. V. Rao,
Physica A {\bf 250}, 327(1998); M. E. Fisher, Physica A {\bf 263}, 554
(1999).

\bibitem{James}  F. James, Comp. Phys. Commun. {\bf 60}, 329 (1990); Chaos,
Solitons Fractals {\bf 6}, 221 (1995).

\bibitem{Yu}  L. Yu, E. Ott, and O. Chen, Phys. Rev. Lett. {\bf 65}, 2935
(1990).

\bibitem{Paladin}  G. Paladin, M. Serva, and A. Vulpiani, Phys. Rev. Lett. 
{\bf 74}, 66 (1995).

\bibitem{Loreto}  V. Loreto, G. Paladin, and A. Vulpiani, Phys. Rev. E {\bf %
53}, 2087 (1996).

\bibitem{Sugihara}  G. Sugihara and R. M. May, Nature (London) {\bf 344}, 734 (1990).

\bibitem{Wales}  D. J. Wales, Nature (London) {\bf 350}, 485 (1991).

\bibitem{Tsonis}  A.A. Tsonis and J. B. Elsner, Nature (London) {\bf 358}, 217 (1992).

\bibitem{Ulam}  S. M. Ulam, {\em A Collection of Mathematical Problems}
(Interscience, New York, 1960).

\bibitem{Stein}  P. Stein and S. M. Ulam, Dissertationes Mathematicae / 
Rozprawy Matematyczne {\bf 39},
401 (1964).

\bibitem{Katsura}  S. Katsura and W. Fukuda, Physica A {\bf 130}, 597 (1985).

\bibitem{Kawamoto}  S. Kawamoto and T. Tsubata, J. Phys. Soc. Jpn. {\bf 65},
3078 (1996).

\bibitem{Brown}  R. Brown and L. O. Chua, Int. J. Bifurcation and Chaos 
Appl. Sci. Eng. {\bf %
6}, 219 (1996).

\bibitem{Umeno}  K. Umeno, Phys. Rev. E {\bf 55}, 5280 (1997).

\bibitem{González}  J. A. Gonz\'{a}lez and L. B. Carvalho, Mod. Phys. Lett.
B {\bf 11}, 521 (1997).

\bibitem{Nazareno}  H. N. Nazareno, J. A. Gonz\'{a}lez, and I. Costa, Phys.
Rev. B {\bf 57}, 13583 (1998).

\bibitem{González2}  J. A. Gonz\'{a}lez and R. Pino, Comput. Phys. Comun. {\bf %
120}, 109 (1999).

\bibitem{Eckmann}  J. P. Eckmann and D. Ruelle, Rev. Mod. Phys. {\bf 57},
617 (1985).

\bibitem{Pincus}  S. Pincus and B. S. Singer, Proc. Natl. Acad.\ Sci. USA 
{\bf 93}, 2083 (1996).

\bibitem{Benzi}  R. Benzi, A. Sutera, and A. Vulpiani, J. Phys. A {\bf 14},
L453 (1981).

\bibitem{McNamara}  B. McNamara and K. Wiesenfeld, Phys. Rev. A {\bf 39},
4859 (1989).

\bibitem{Bulsara}  A. R. Bulsara and L. Gammaitoni, Phys. Today {\bf 49},
39 (1996).

\bibitem{Gammaitoni}  L. Gammaitoni, P. H\"{a}nggi, P. Jung, and F.
Marchesoni, Rev. Mod. Phys. {\bf 70}, 223 (1998).

\bibitem{Wiesenfeld}  K. Wiesenfeld and F. Moss, Nature (London) {\bf 373}, 33 (1995).

\bibitem{Ippen}  E. Ippen, J. Lindne, and W. Ditto, J. Stat. Phys. {\bf 70},
148 (1993).

\bibitem{Carrol}  T. L. Carrol and L. M. Pecora, Phys. Rev. Lett. {\bf 70},
576 (1993).

\bibitem{Kapitaniak}  T. Kapitaniak, Phys. Rev. E {\bf 49}, 5855 (1994).

\bibitem{Yang}  W. Yang, M. Ding, and G. Hu, Phys. Rev. Lett. {\bf 74}, 3955
(1995).

\bibitem{Crisanti}  A. Crisanti, M. Falcioni, G. Paladin, and A. Vulpiani,
J. Phys. A {\bf 27}, L597 (1994).

\bibitem{Anishchenko}  V. S. Anishchenko, A. B. Neimann, and M.\ A.
Safonova, J. Stat. Phys. {\bf 70}, 183 (1993).

\bibitem{Hanggi}  P. H\"{a}nggi, P. Jung, C. Zerbe, and F. Moss, J. Stat.
Phys. {\bf 70}, 25 (1993).

\bibitem{Gammaitoni2}  L. Gammaitoni, Phys. Rev. E {\bf 52}, 4691 (1995).

\bibitem{Gingl}  Z. Gingl, L. B. Kiss, and F. Moss, Europhys. Lett. {\bf 29}%
, 191 (1995).

\bibitem{Chapeau-Blondeau}  F. Chapeau-Blondeau and K. Godivier, Phys. Rev.
E {\bf 55}, 1478 (1997).

\bibitem{Bezrukov}  S. M. Bezrukov and I. Vodyanoy, Nature (London) {\bf 385}, 319
(1997).

\bibitem{Vilar}  J. M. G. Vilar, G. Gomila, and J. M. Rub\'{\i }, Phys. Rev.
Lett. {\bf 81}, 14 (1998).

\bibitem{Marchesoni}  F. Marchesoni, L. Gammaitoni, and A. R. Bulsara, Phys.
Rev. Lett. {\bf 76}, 2609 (1996).

\bibitem{González3}  J. A. Gonz\'{a}lez, B. A. Mello, L. I. Reyes, and L. E.
Guerrero, Phys. Rev. Lett. {\bf 80}, 1361 (1998).

\bibitem{Witt}  A. Witt, A. Neiman, and J. Kurth, Phys. Rev. E {\bf 55},
5050 (1997).

\bibitem{González4}  J. A. Gonz\'{a}lez and J. A. Ho\l yst, Phys. Rev. B 
{\bf 45}, 10338 (1992).

\bibitem{González5}  J. A. Gonz\'{a}lez and B. A. Mello, Phys. Scr. {\bf 54}%
, 14 (1996).

\bibitem{González6}  J. A. Gonz\'{a}lez, L. E. Guerrero, and A. Bellor\'{\i %
}n, Phys. Rev. E {\bf 54}, 1265 (1996).

\bibitem{González7}  J. A. Gonz\'{a}lez, A. Bellor\'{\i }n, and L. E.
Guerrero, Phys. Rev. E {\bf 60}, R37 (1999).

\bibitem{Chua}  L. O. Chua, C. A. Desoer, and E. S. Kuh, {\em Linear and
Nonlinear Circuits} (McGraw-Hill, New York, 1987).

\bibitem{Chacón}  R. Chac\'{o}n, Phys. Rev. Lett. {\bf 77}, 482 (1996).

\bibitem{Hartle}  J. B. Hartle, Complexity {\bf 3}, 22 (1997).

\bibitem{Stone}  L. Stone, P. I. Saparin, A. Huppert, and C. Price, Geophys.
Res. Lett. {\bf 25}, 175 (1998).

\bibitem{Saltzman}  B. Saltzman, H. Hu, and R. J. Oglesby, Dyn. 
Atmos. Oceans {\bf 27}, 619 (1997).

\bibitem{Liu}  H-S. Liu and B. F. Chao, J. Atmos. Sci. {\bf 55}, 227 (1998).

\bibitem{Babinec}  P. Babinec, Phys. Lett A {\bf 225}, 179 (1997).
\end{references}
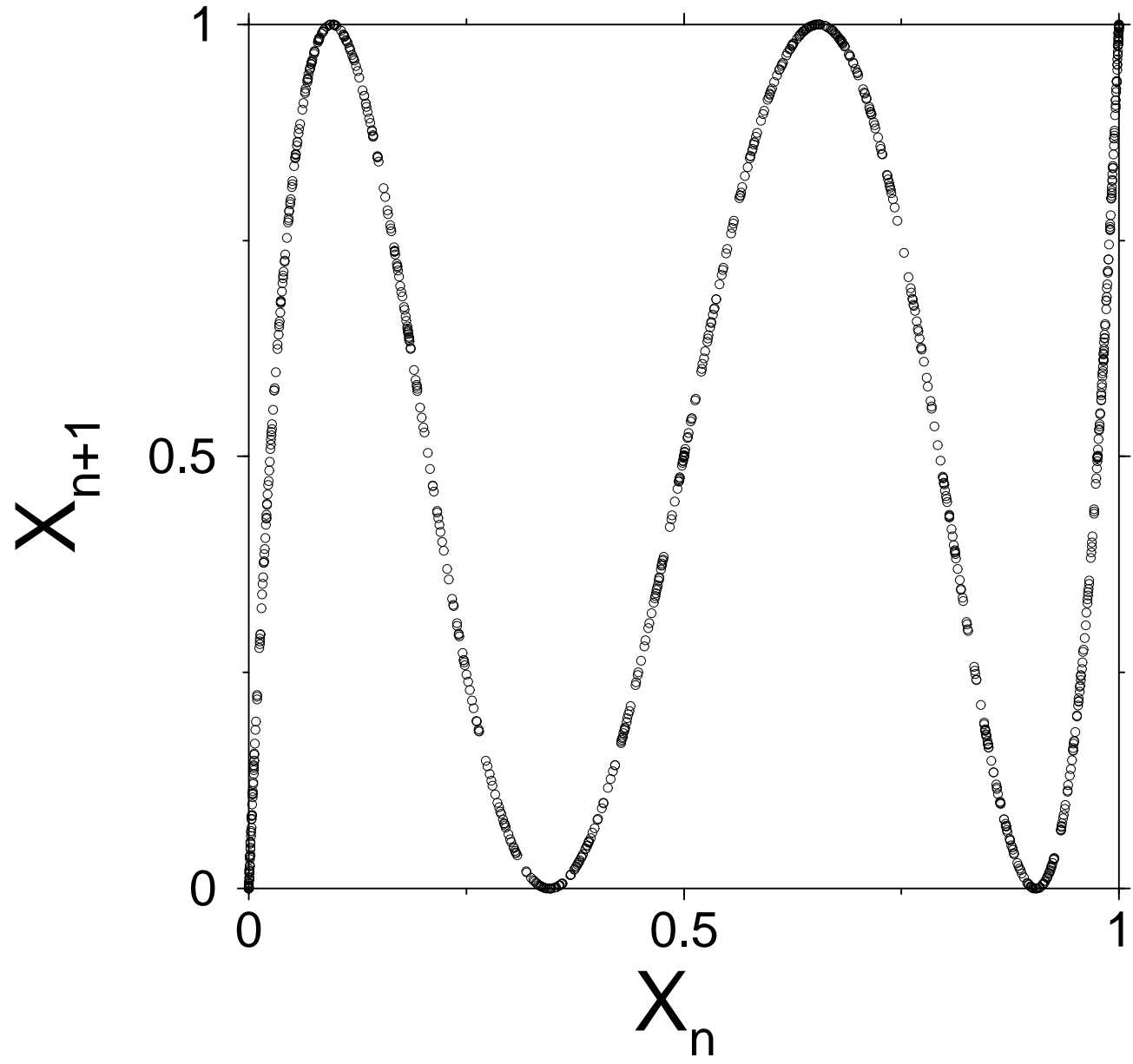
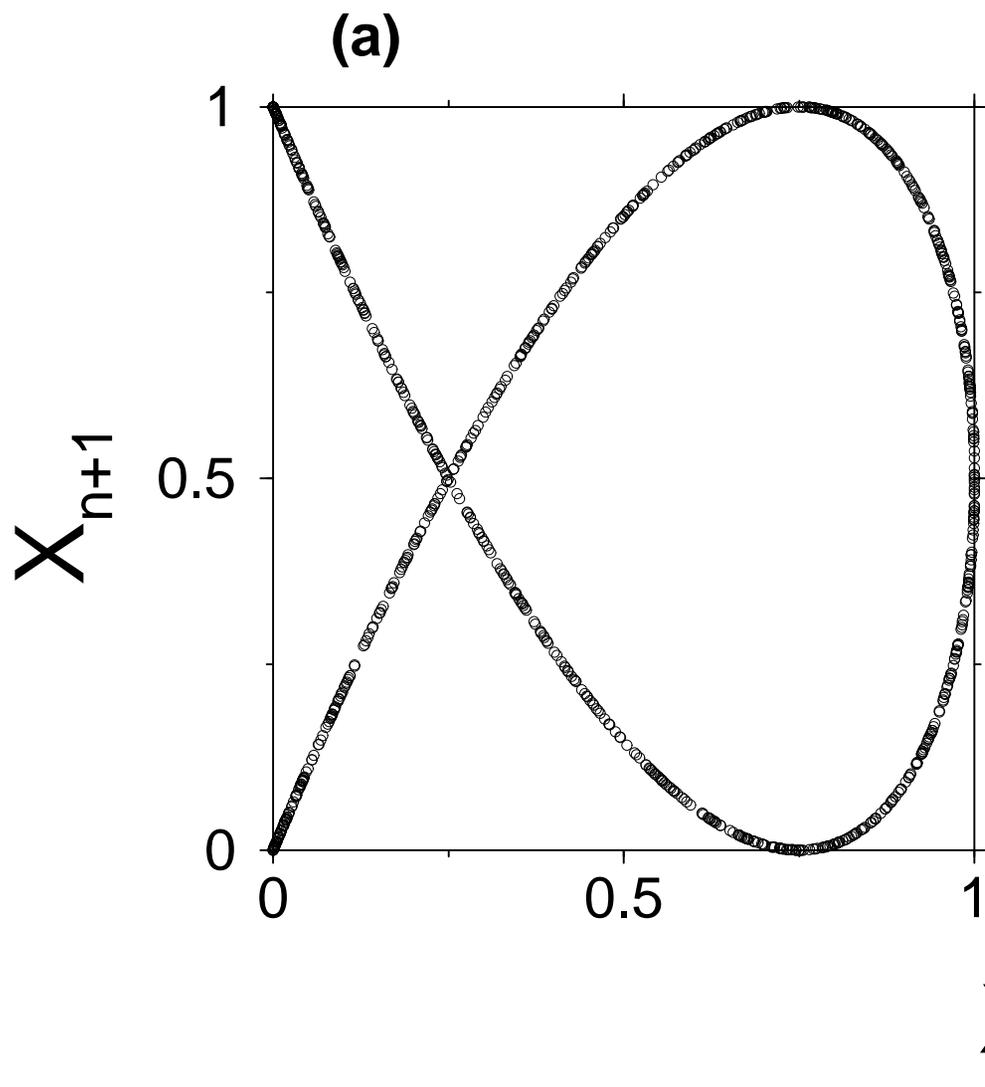
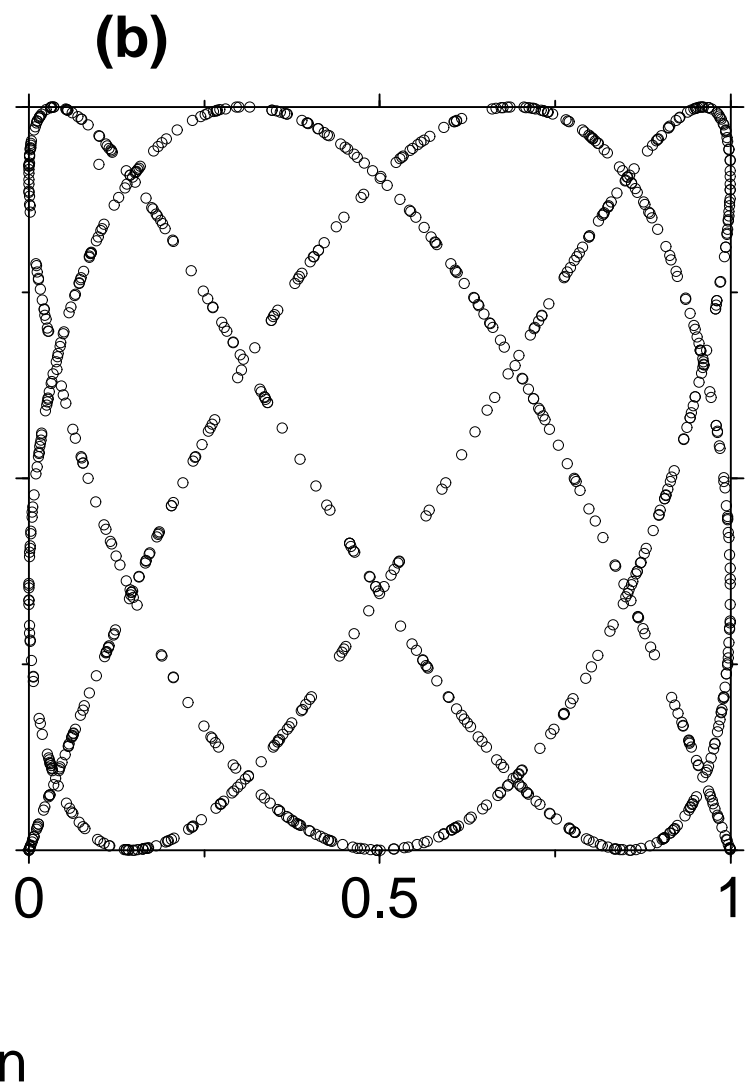
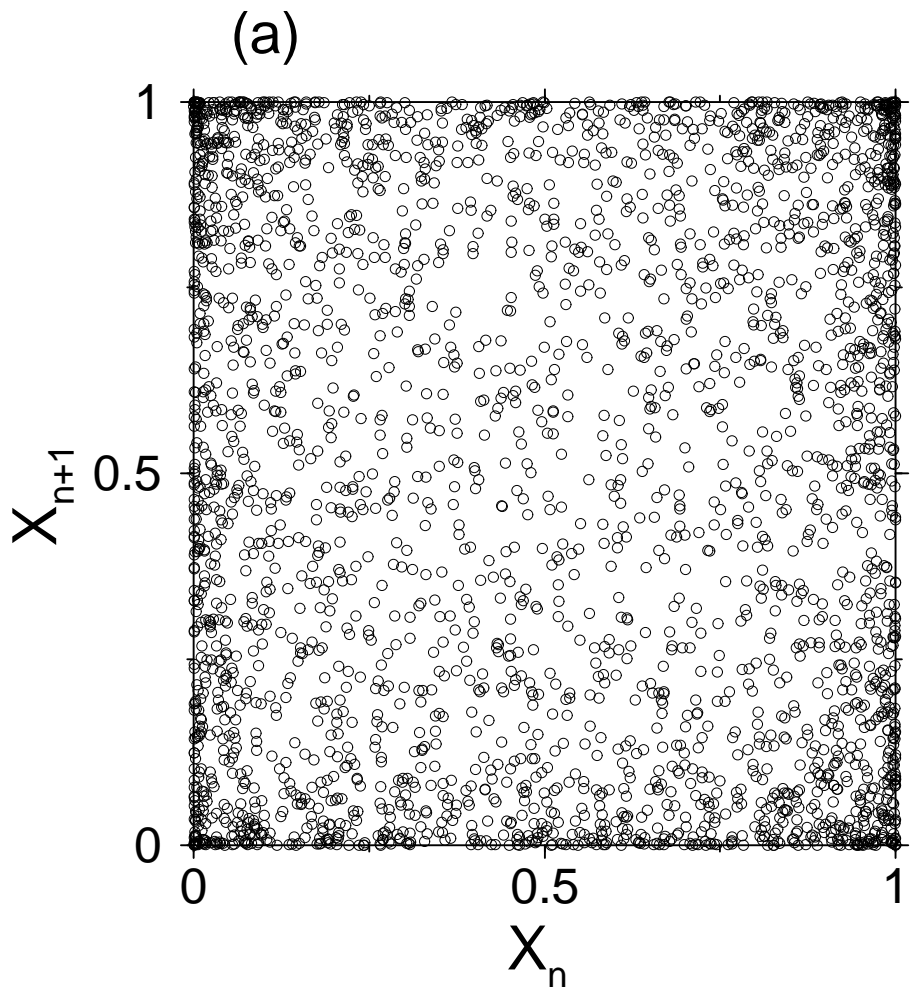
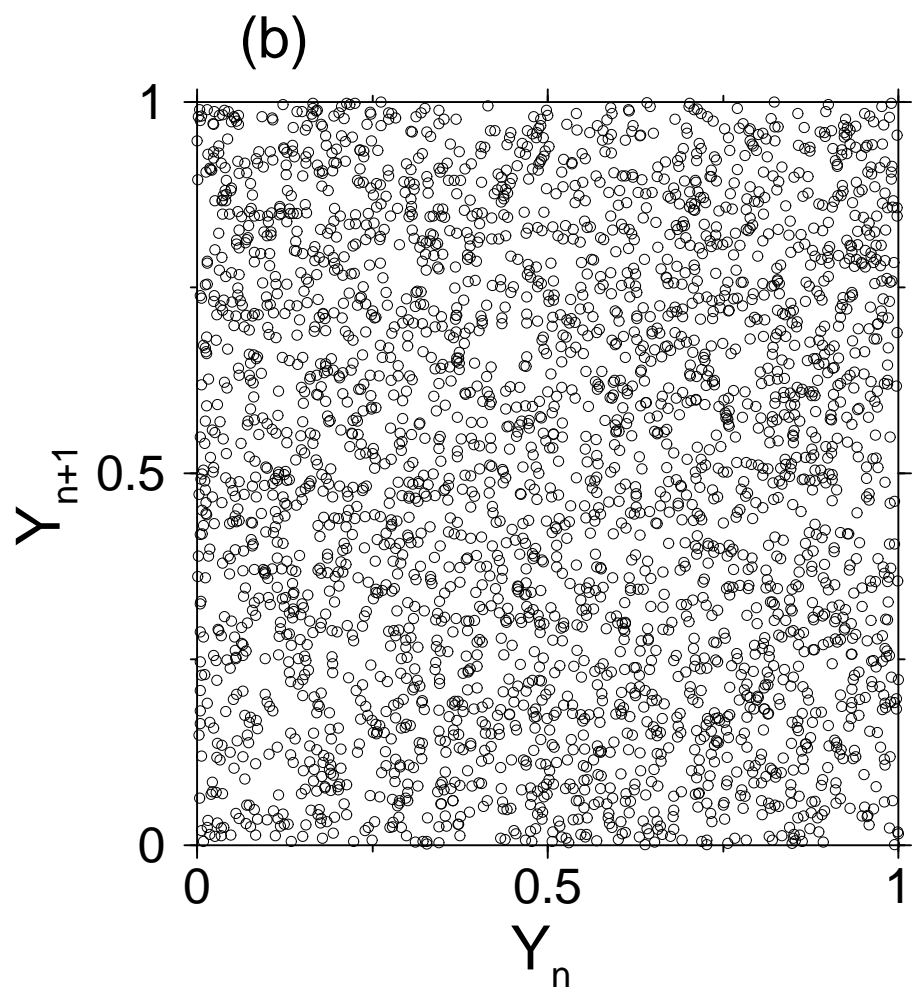
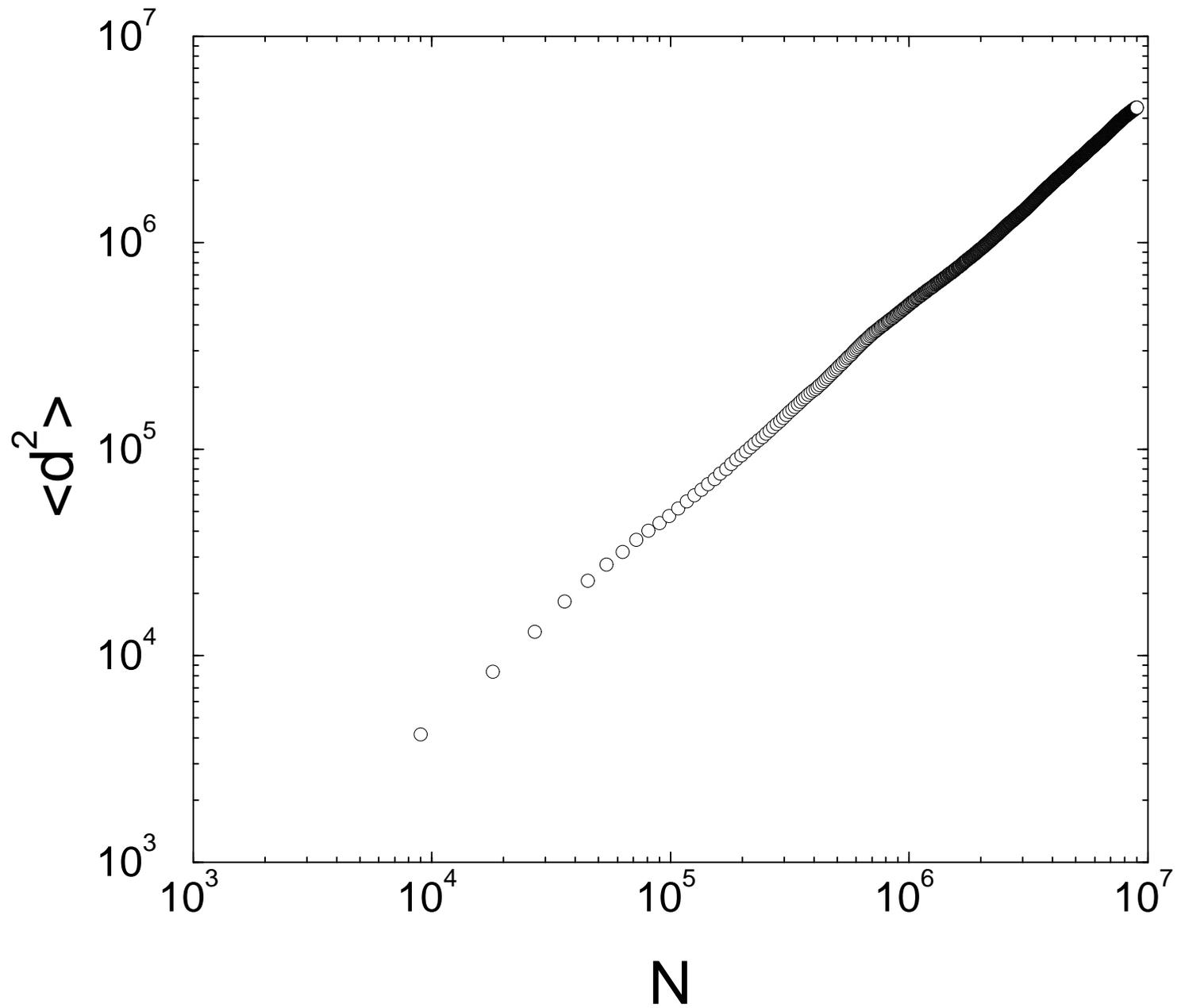
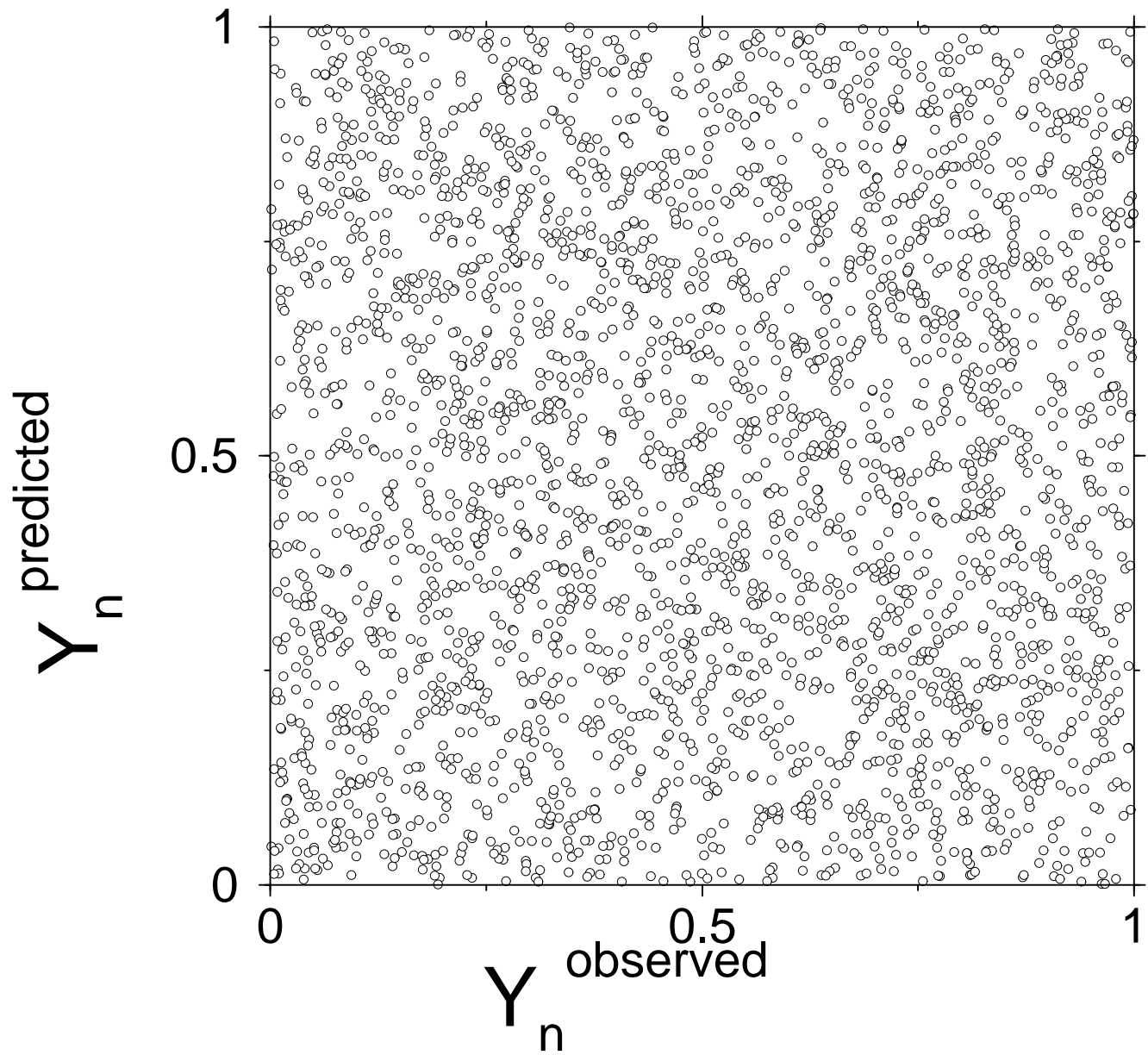
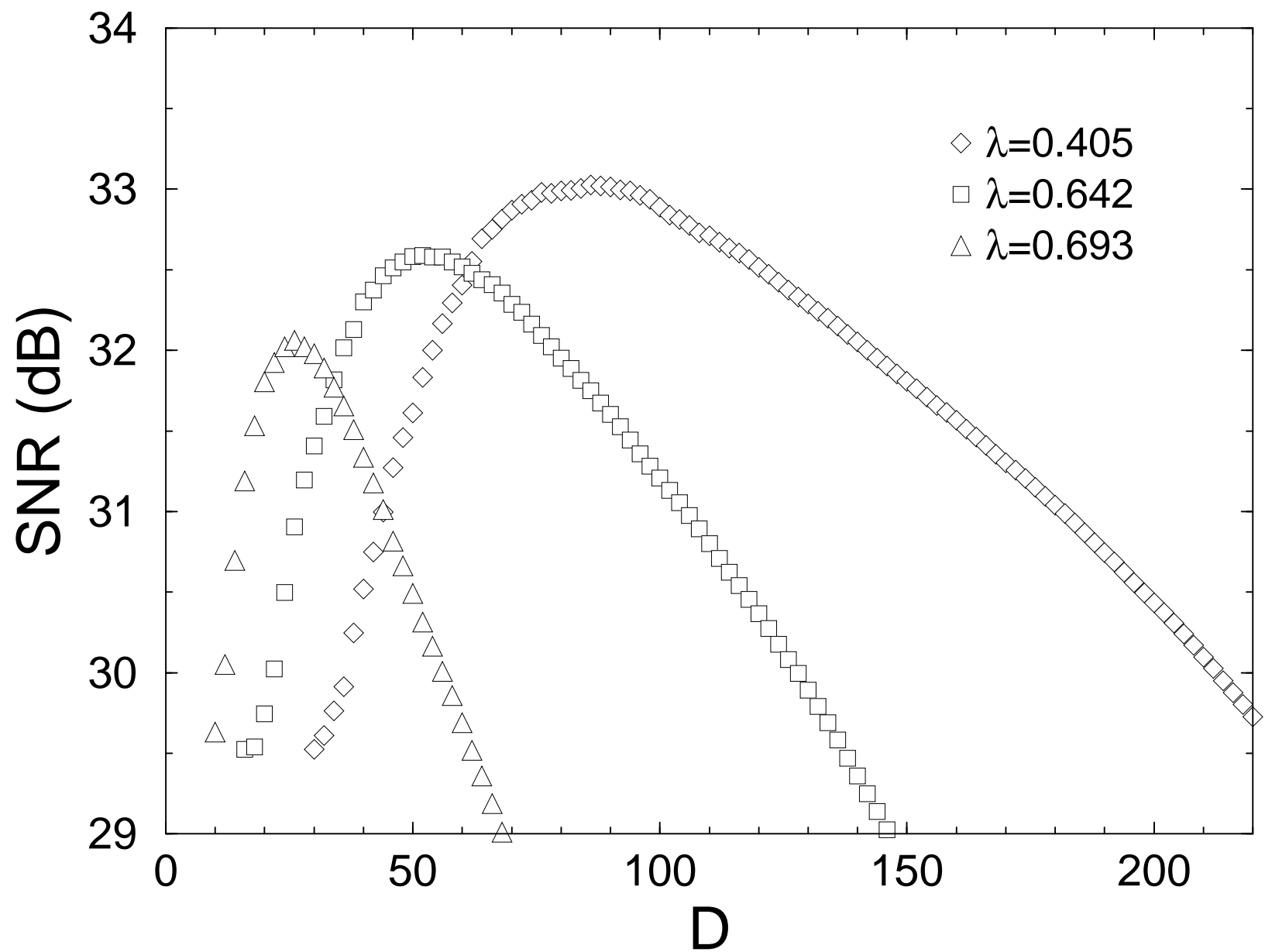
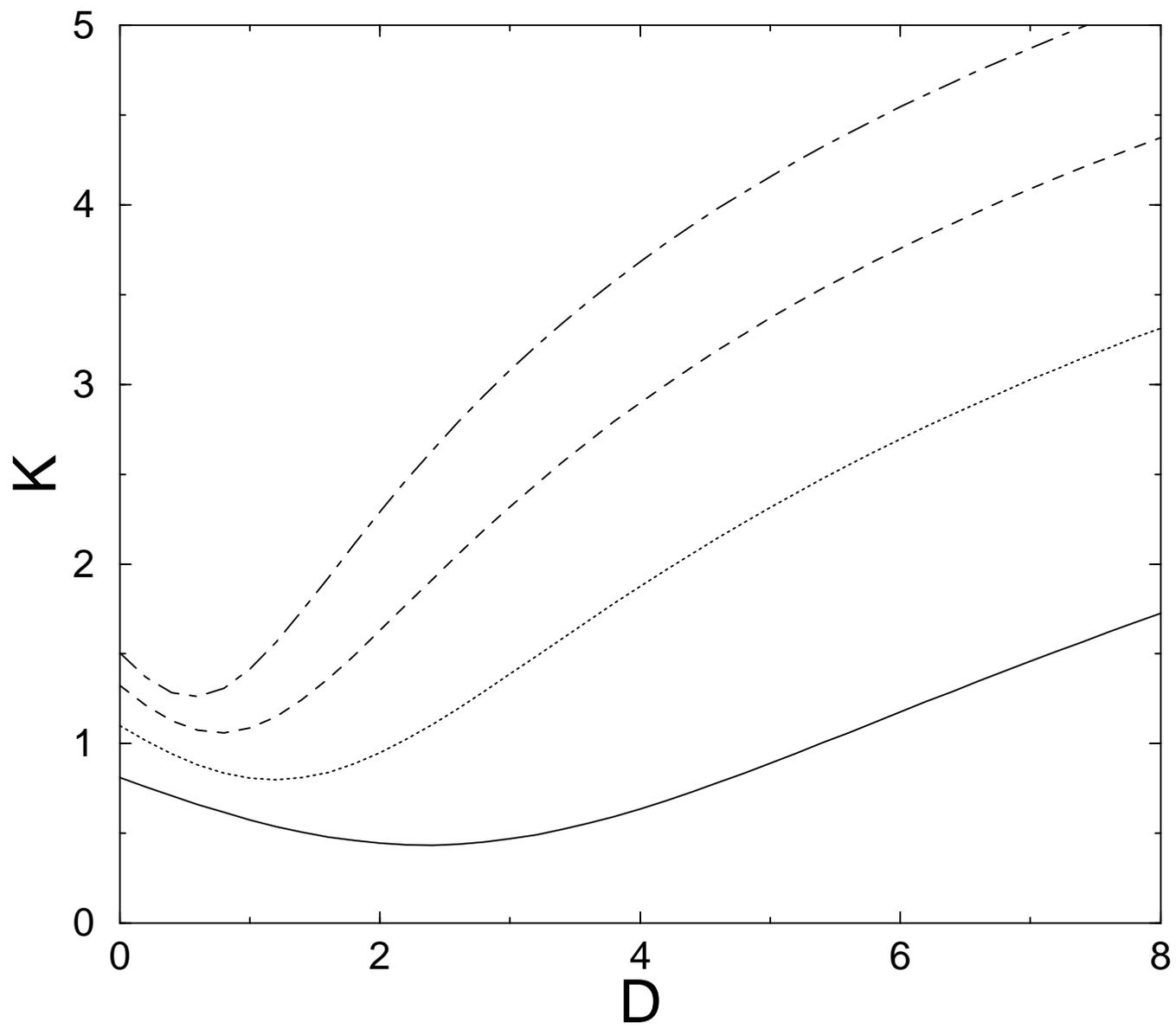
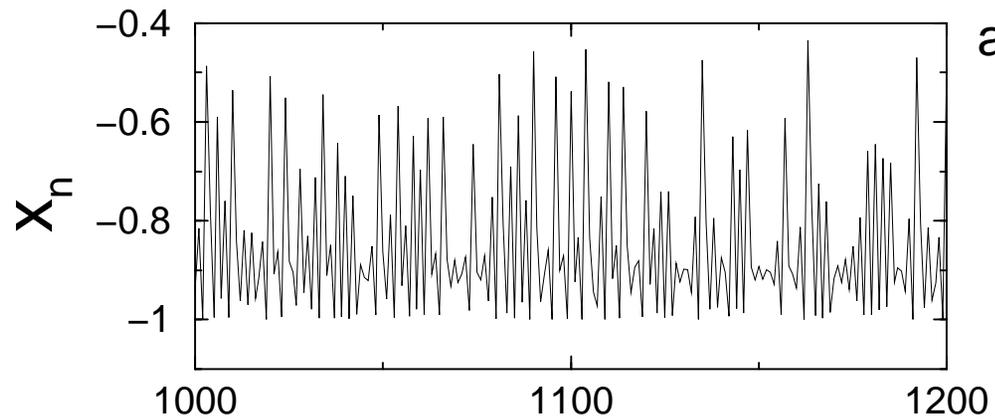
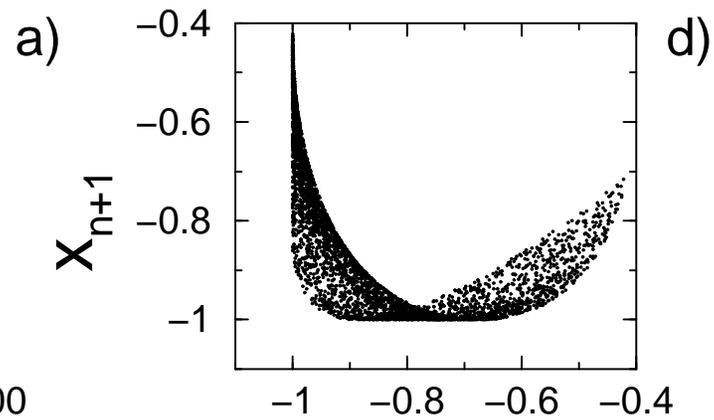
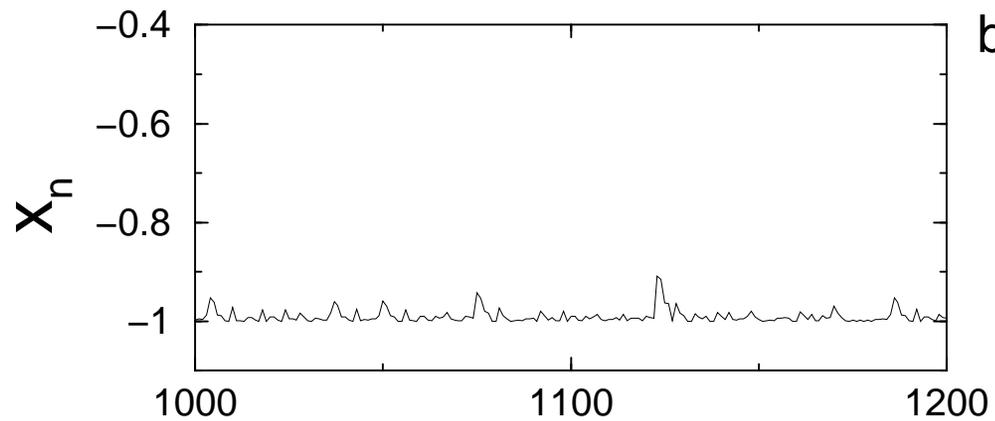
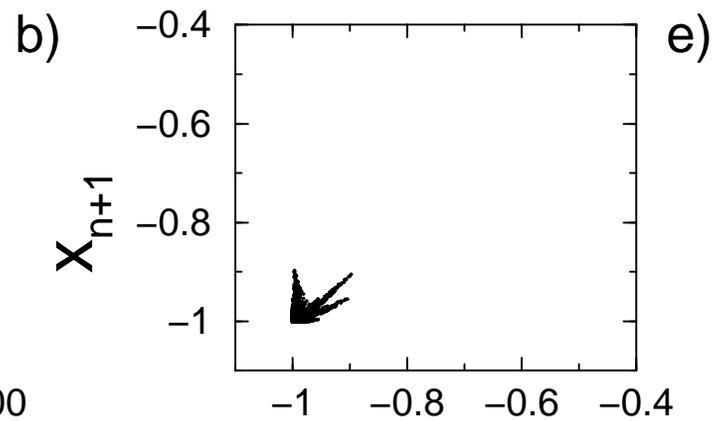
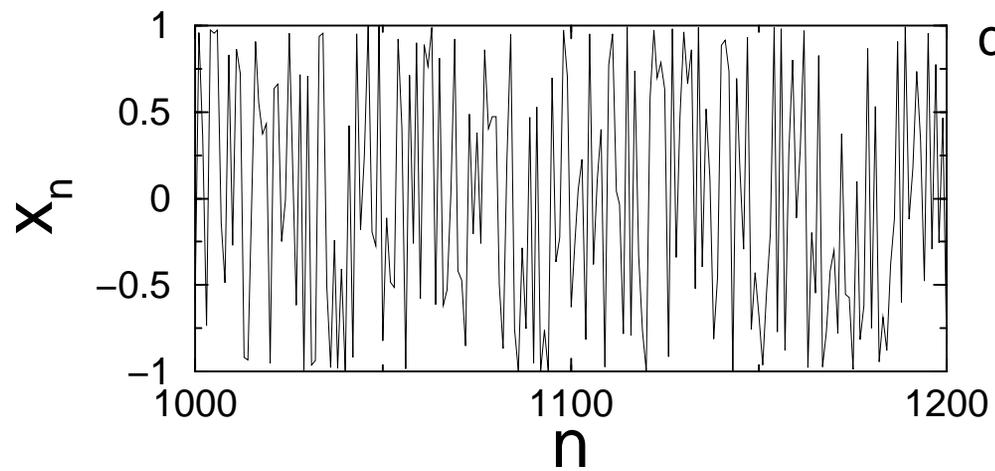
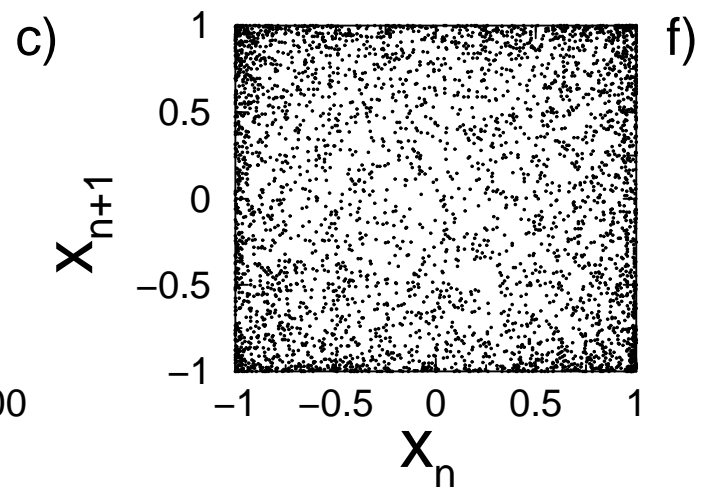
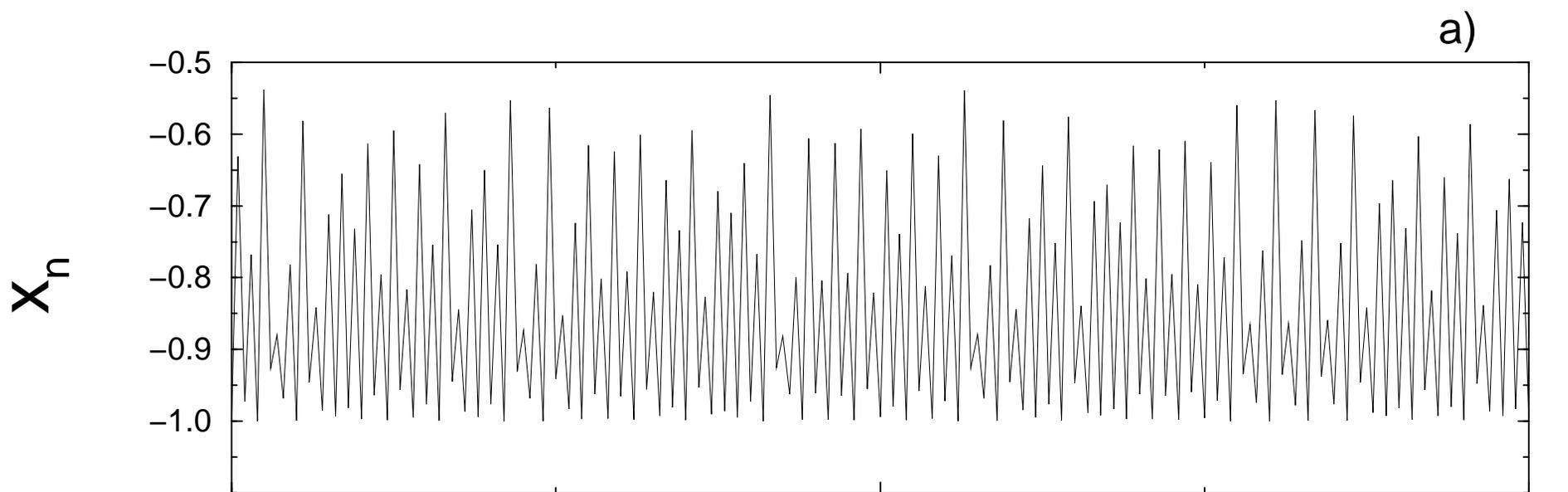
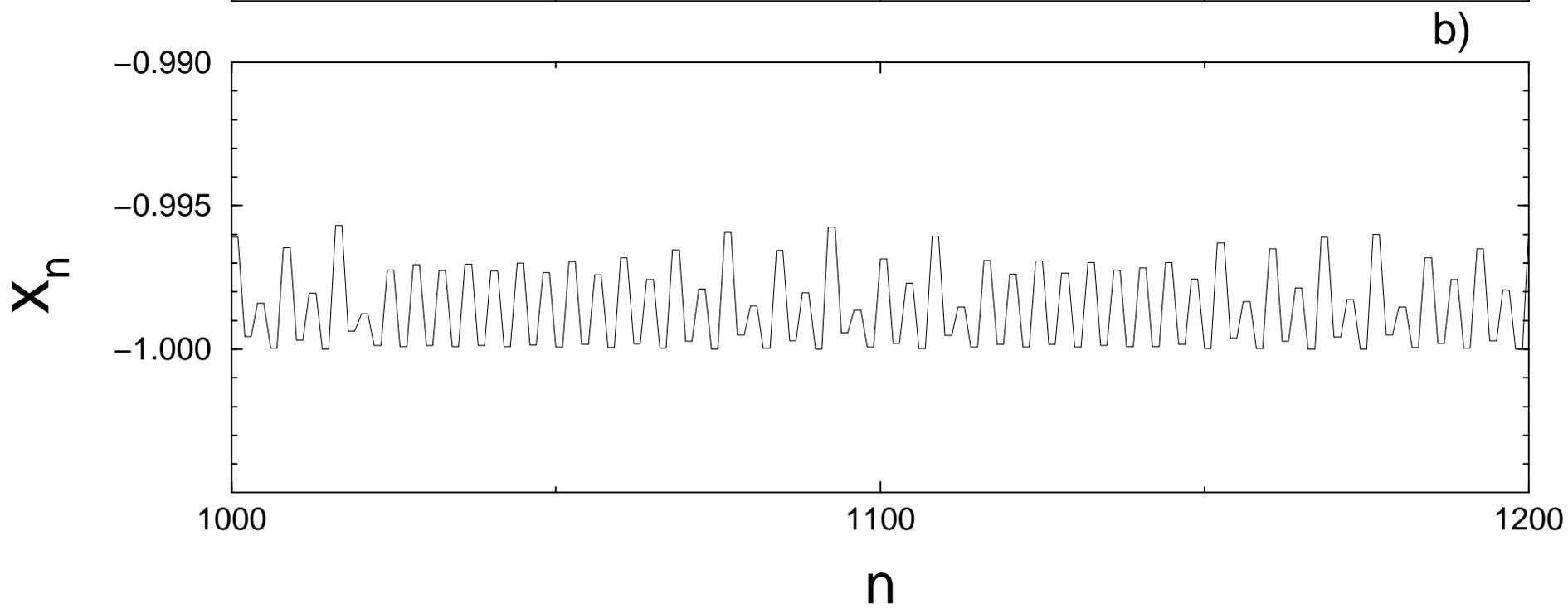
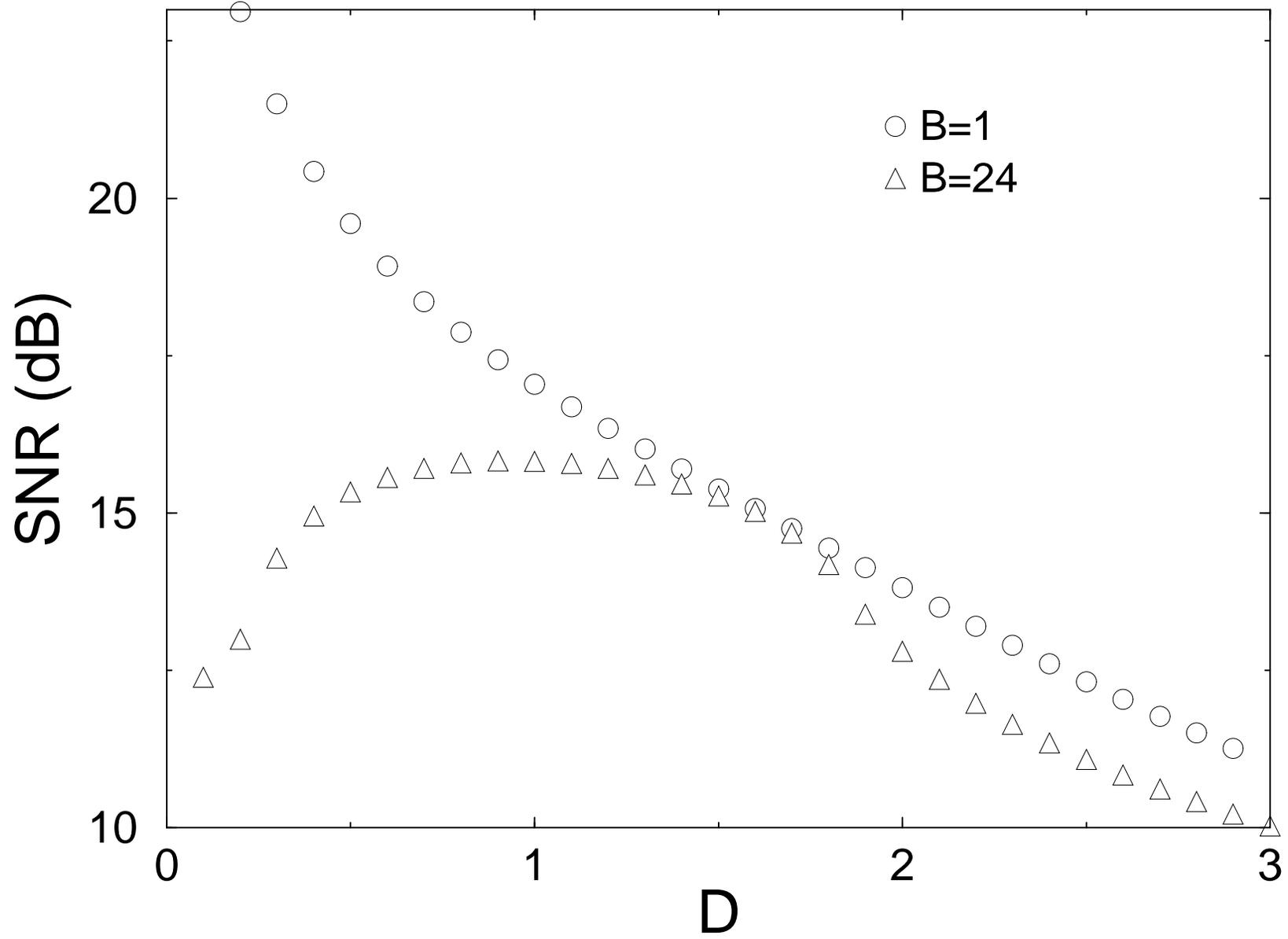
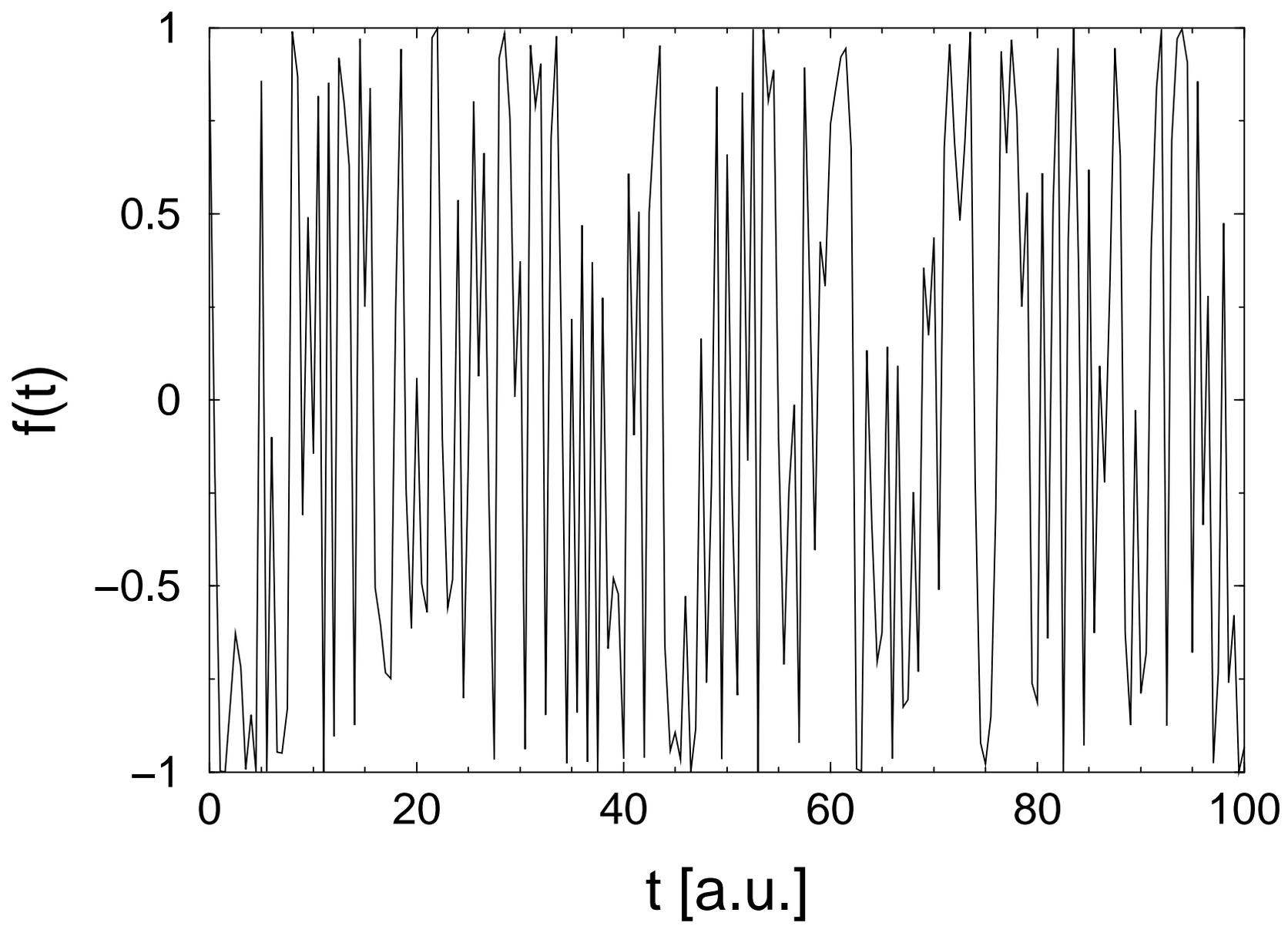
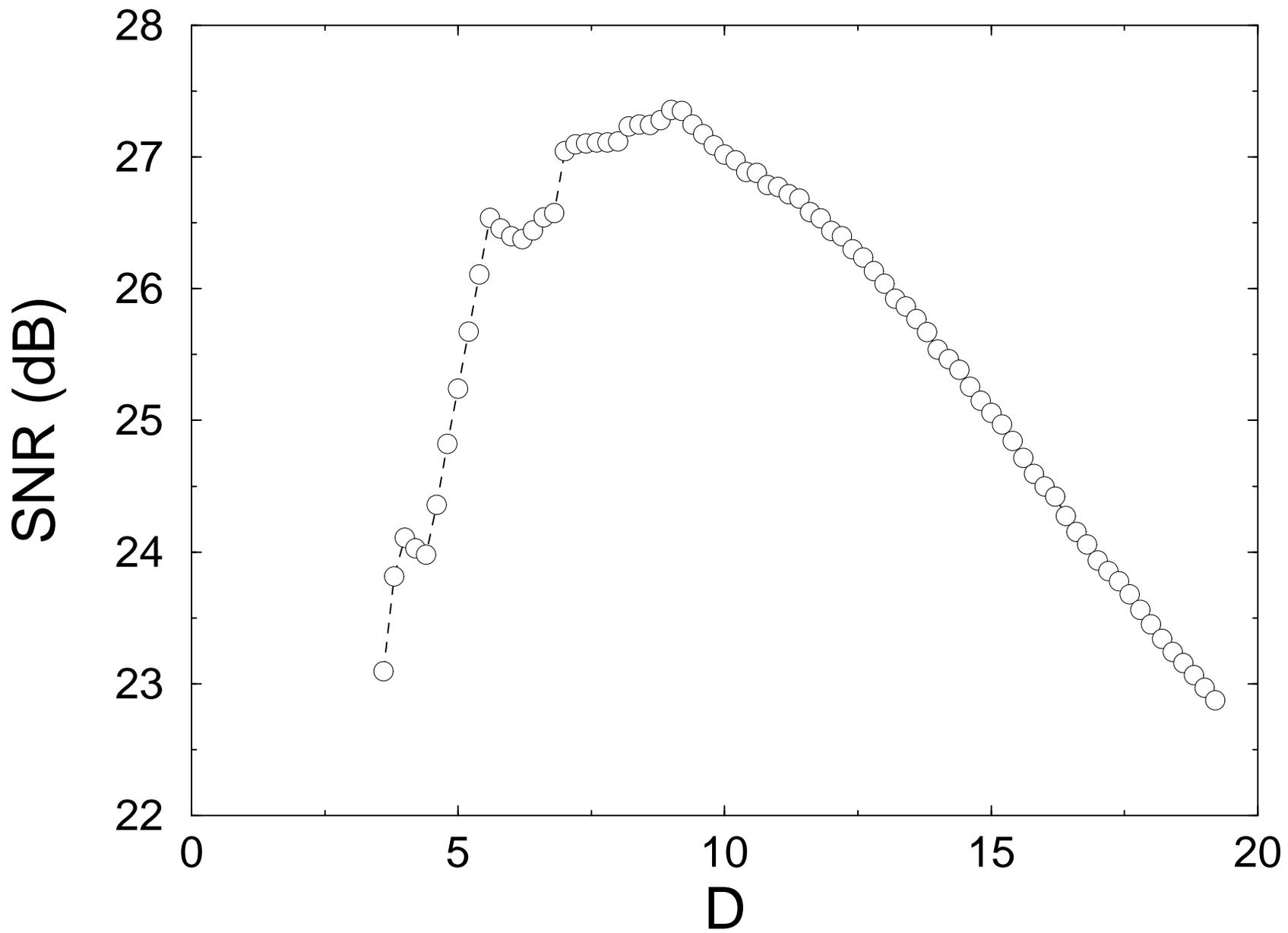
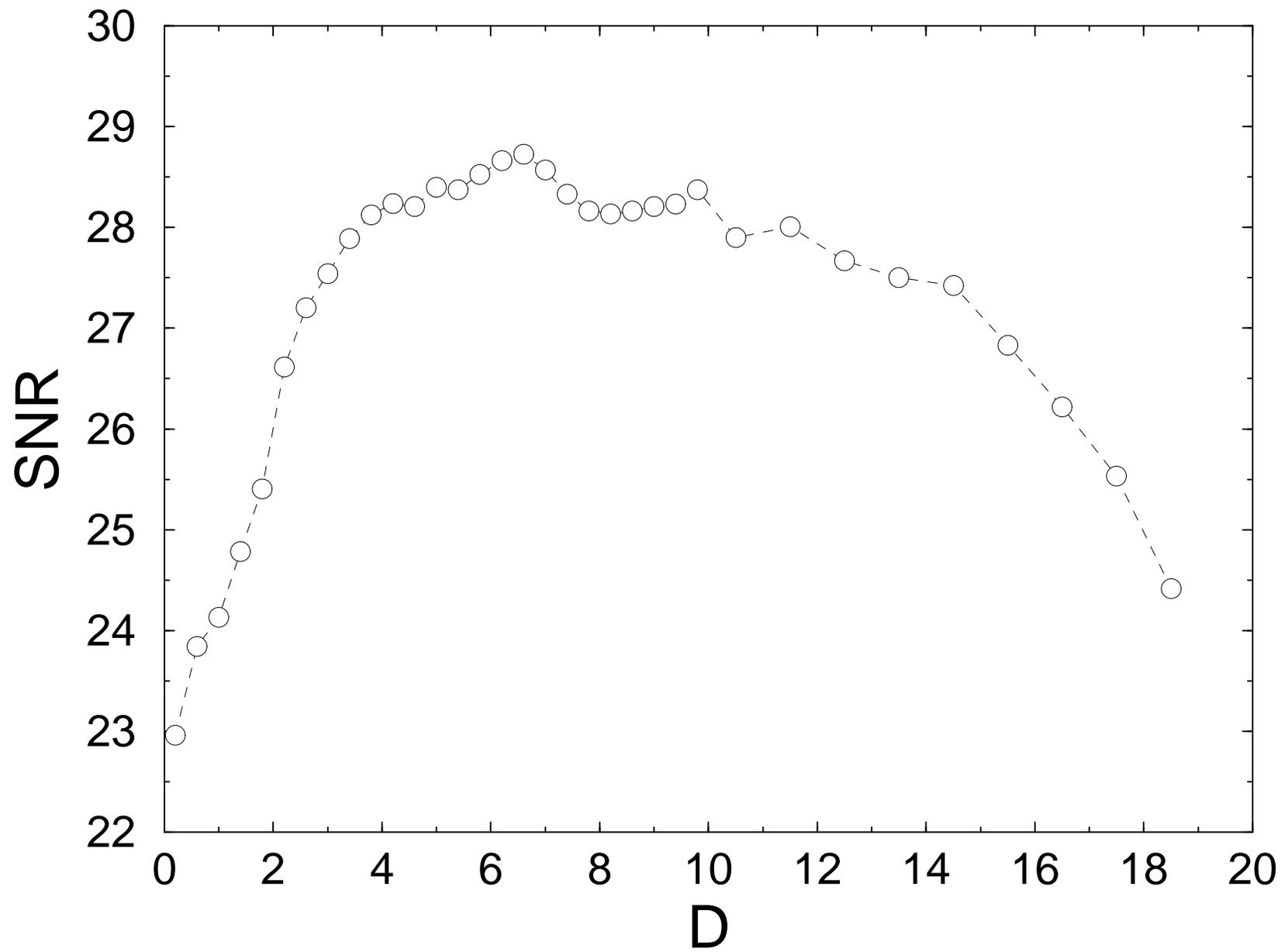
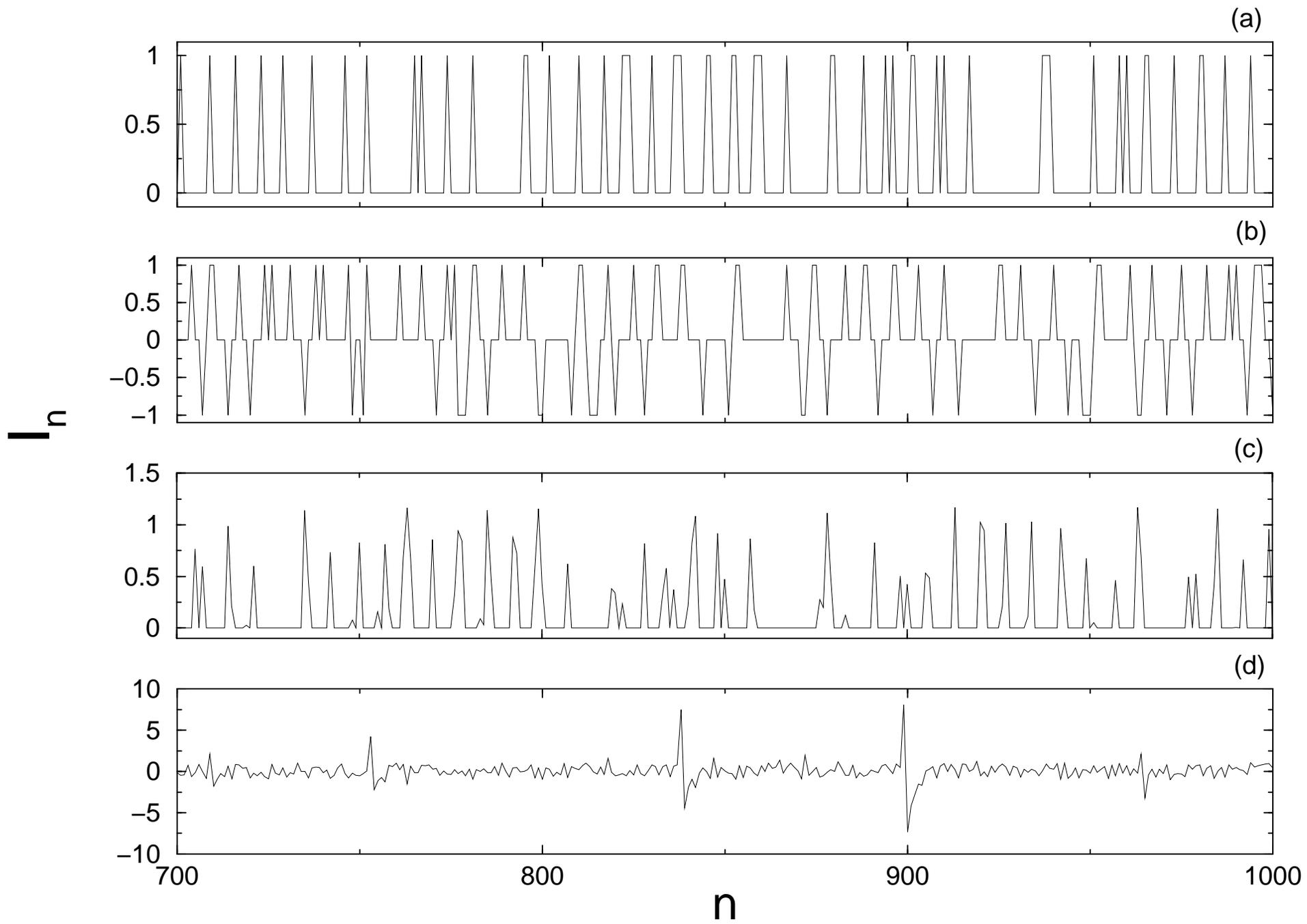
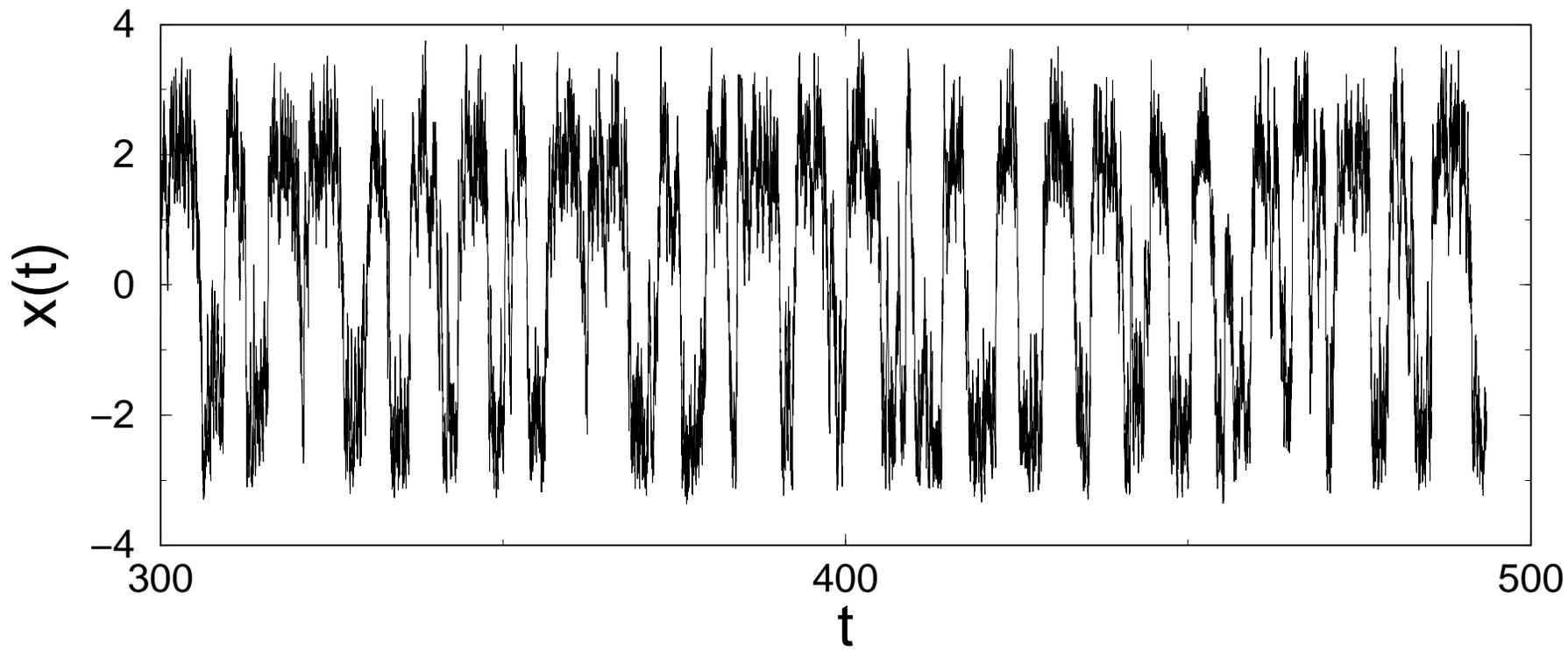
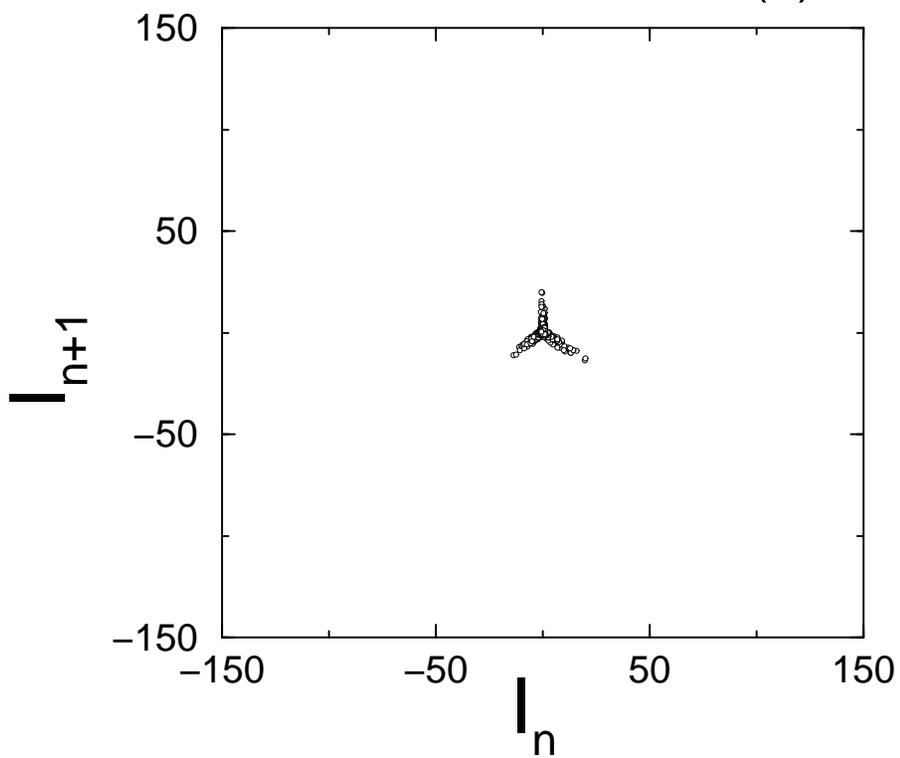
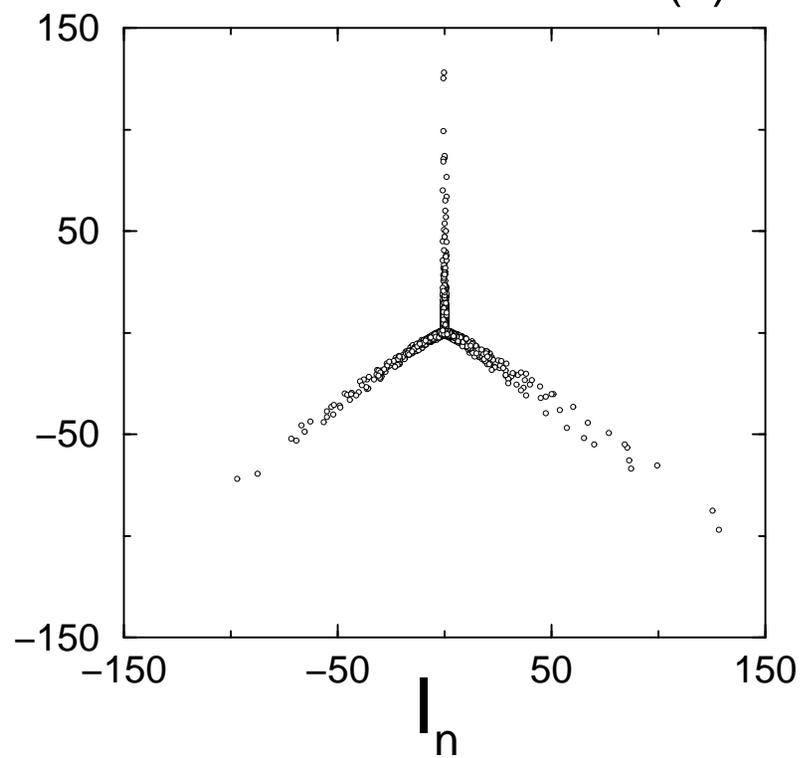
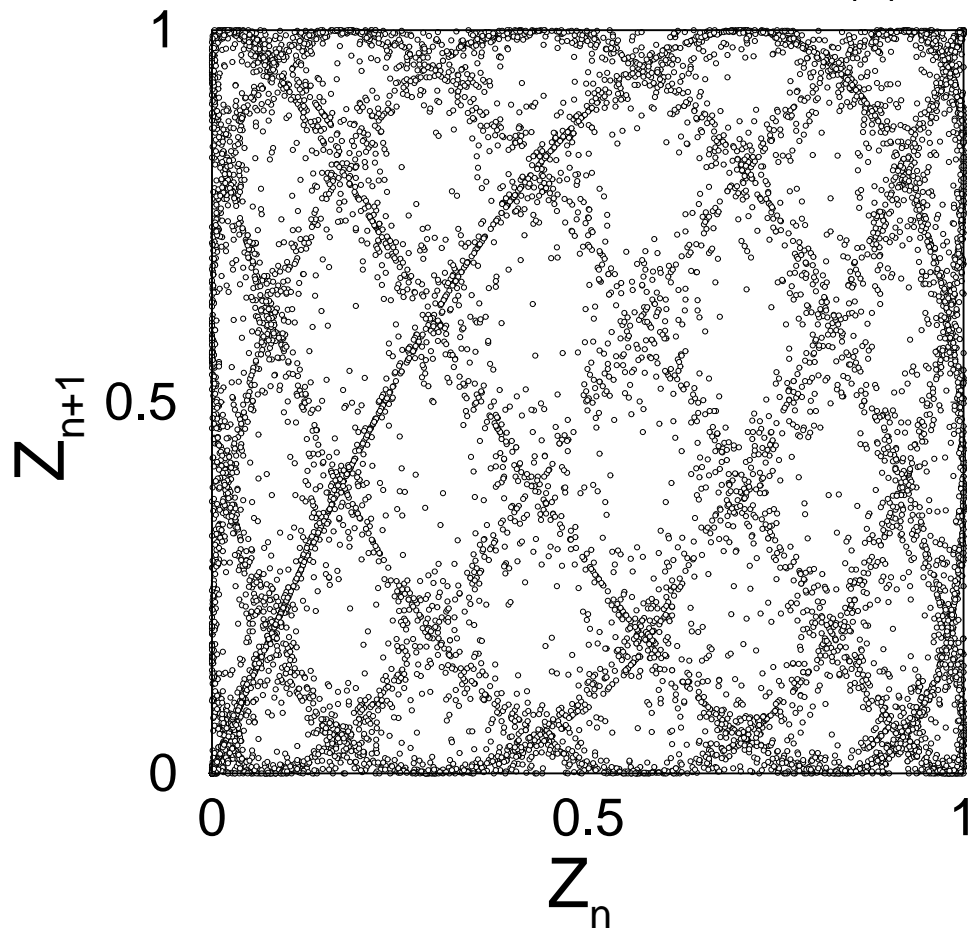
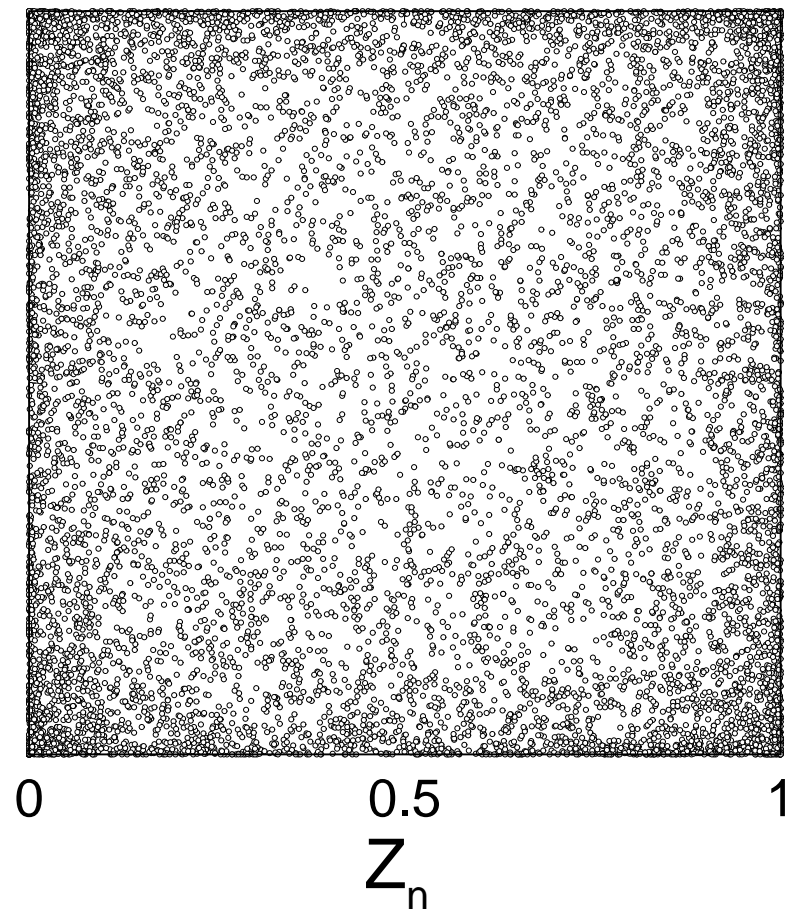
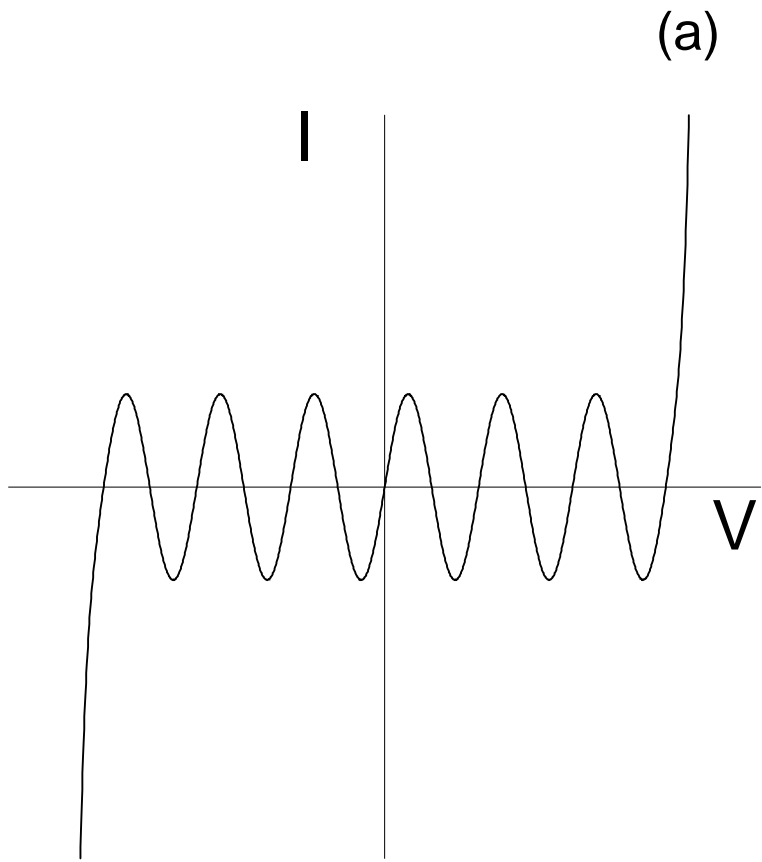
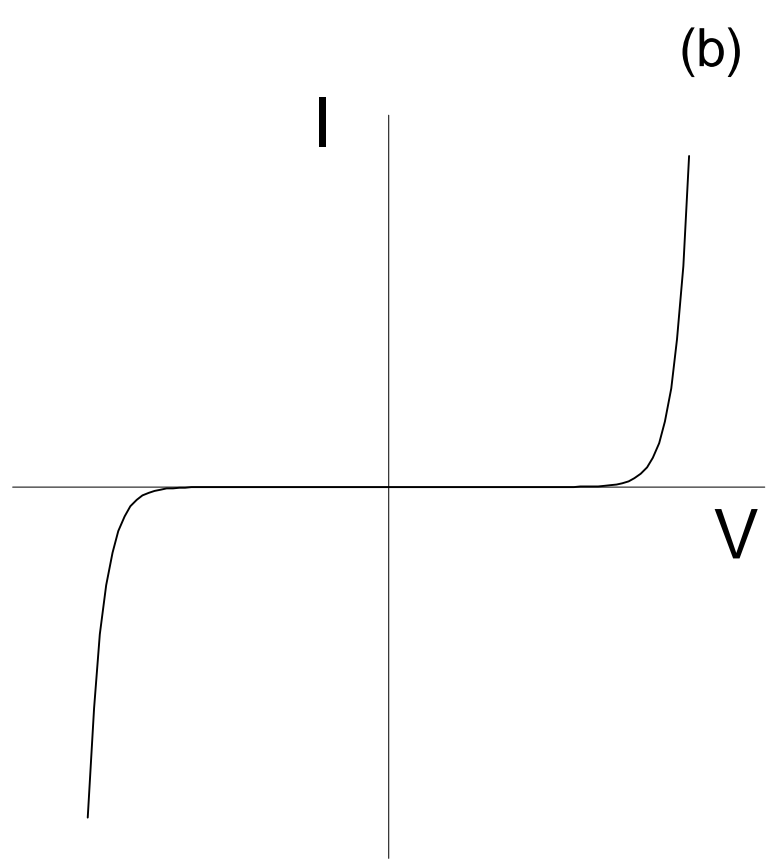
\end{document}